\RequirePackage{fix-cm}
\documentclass[twocolumn]{svjour3}
\smartqed
\usepackage{graphicx}
\usepackage{amsmath}

\usepackage{makecell}
\usepackage{xcolor, colortbl}
\usepackage{lipsum}
\usepackage{multirow}

\usepackage{subcaption}
\usepackage{pdfpages}
\usepackage{amssymb}

\usepackage{bbding}

\usepackage{fontawesome}

\newcommand{\eg}{e.\,g.,\ }
\newcommand{\ie}{i.\,e.,\ }

\newcommand{\labeltitle}[1]{\vskip 0.03in \noindent\textbf{#1}} 

\usepackage{csquotes}

\usepackage{listings}

\usepackage{xurl} 
\usepackage{microtype}

\NewDocumentCommand\DownArrow{O{2.0ex} O{black}}{%
   \mathrel{\tikz[baseline] \draw [<-, line width=0.5pt, #2] (0,0) -- ++(0,#1);}
}

\NewDocumentCommand\UpArrow{O{2.0ex} O{black}}{%
   \mathrel{\tikz[baseline] \draw [->, line width=0.5pt, #2] (0,0) -- ++(0,#1);}
}

\NewDocumentCommand\NoArrow{O{2.0ex} O{black}}{%
   \mathrel{\tikz[baseline] \draw [-, line width=0.5pt, #2] (0,0.15) -- ++(#1,0);}
}

\definecolor{darkgreen}{RGB}{50,113,7}
\definecolor{darkred}{RGB}{182, 2, 3}
\definecolor{darkorange}{RGB}{236, 112, 27}

\NewDocumentCommand\Yes{O{2.0ex} O{black}}{%
   \textcolor{green}{\textbf{O}}
}

\NewDocumentCommand\greenCheck{O{2.0ex} O{black}}{%
   \textcolor{darkgreen}{\faThumbsOUp}
}

\NewDocumentCommand\orangeCheck{O{2.0ex} O{black}}{%
   \textcolor{darkorange}{\faGears}
}

\NewDocumentCommand\No{O{2.0ex} O{black}}{%
   \textcolor{darkred}{\faThumbsODown}
}

\begin{document}

\title{LakeVilla: A Modular and Non-Invasive Toolbox for Lakehouse Transactions}

\author{Tobias G\"otz       \and
        Daniel Ritter \and
        Jana Giceva
}

\institute{Tobias G\"otz \at
              Technical University of Munich \\
              Munich, Germany\\
              \email{goetzt@in.tum.de}  \\
              ORCID: 0009-0009-5557-5518
           \and
           Daniel Ritter \at
              SAP SE \\
              Walldorf, Germany\\
              \email{daniel.ritter@sap.com}\\
              ORCID: 0000-0001-6146-3365
           \and 
           Jana Giceva \at
              Technical University of Munich \\
              Munich, Germany\\
              \email{jana.giceva@in.tum.de}   \\
              ORCID: 0000-0002-1926-3551     
}

\date{Received: date / Accepted: date}

\maketitle

\begin{abstract}
Data lakehouses (LHs) are at the core of current cloud analytics stacks by providing elastic, relational compute on data in cloud data lakes across vendors.
For relational semantics, they rely on open table formats (OTFs). Unfortunately, they have many missing features inherent to their metadata designs, like no support for multi-table transactions and recovery in case of an abort in concurrent, multi-query workloads. This, in turn, can lead to non-repeatable reads, stale data, and high costs in production cloud systems. 
In this work, we introduce LakeVilla, a modular toolbox that introduces recovery, complex transactions, and transaction isolation to state-of-the-art OTFs like Apache Iceberg and Delta Lake tables. We investigate its transactional guarantees and show it has minimal impact on performance (2\% YCSB writes, 2.5\% TPC-DS reads) and provides concurrency control for multiple readers and writers for arbitrary long transactions in OTFs in a non-invasive way.
\end{abstract}

\section{Introduction}

\begin{table*}[t]
\small
\caption{Existing approaches with their supported features and the respective design area}
\centering
\label{fig:overview}
\begin{tabular}{l|c|ccc|cc}
&  & \multicolumn{3}{c|}{\textbf{Multi-Query/Table: (i) -- (iii)}} & \multicolumn{2}{c}{\textbf{Design Decisions}}\\\hline
  \makecell[ll]{System/Extension} &  \makecell[cc]{Single-\\query Txn} & \makecell[cc]{Reco-\\very (i)} & \makecell[cc]{Complex\\Txn (ii)}& \makecell[cc]{Isola-\\tion (iii)} &  \makecell[cc]{Built into the\\OTF structure (D1)} & \makecell[cc]{Modular solution for\\OTF transactions (D2)}  \\ \hline
    Apache Iceberg \cite{Iceberg_website} & \greenCheck & \No & \orangeCheck & \No & \No & \No  \\
        Delta Lake \cite{journals/pvldb/ArmbrustDPXZ0YM20,delta4_multi_table} & \greenCheck & \No & \orangeCheck & \orangeCheck  & \No & \No  \\
        Apache Hudi \cite{hudi_concurrency_docs} & \greenCheck & \No & \No & \No & \No & \No   \\
        \hline
        Project Nessie \cite{project_nessie_web} & \greenCheck & \No & \orangeCheck & \orangeCheck & \No & \No   \\
        XTable \cite{xtable_web} & \greenCheck & \No & \No & \No  & \No  & \No  \\
        DuckLake \cite{ducklake} & \greenCheck & \textbf{?} & \greenCheck & \greenCheck & \No & \No \\
        \hline
        LakeVilla (this paper)& \greenCheck & \greenCheck & \greenCheck & \greenCheck & \greenCheck & \greenCheck  \\
\end{tabular}

\small
Provided: \greenCheck \quad Limited/Planned: \orangeCheck \quad Unknown due to technical limitations (Section \ref{sec:related_work}): \textbf{?} \quad Not provided: \No
\end{table*}

\labeltitle{Lakehouses for analytical workloads.} The past few years have seen the rise of a new type of data management system, called lakehouse (LH), which combines the benefits of low-cost, open-format data lakes and transactional data warehouses~\cite{journals/jpdc/ErramiHKB23,journals/pvldb/Kraft0ZBSYZ23,conf/sigmod/LevandoskiCDDEH24,conf/cidr/Zaharia0XA21}.
With the data gravity leading towards data lakes, LHs for analytical query processing became crucial for various tasks~\cite{redshift_dl_support,conf/sigmod/ArmenatzoglouBB22,conf/bigdataconf/BegoliGK21,journals/pvldb/HaasL23,conf/cidr/Hambardzumyan23,gartner_cloud,journals/pacmmod/KuschewskiSAL23,conf/sigmod/LevandoskiCDDEH24,DBLP:conf/sigmod/SchmidtKHSK24,conf/cidr/Zaharia0XA21}. LHs use open table formats~(OTFs) such as Delta Lake~\cite{delta_lake_web}, Apache Iceberg~\cite{Iceberg_website}, or Apache Hudi \cite{hudi_web} to define separate metadata layers on the object store for each table \cite{journals/pvldb/ArmbrustDPXZ0YM20,conf/cidr/Zaharia0XA21}. Those formats implement transactions for single tables and operations, indexes, and time travel \cite{journals/pvldb/ArmbrustDPXZ0YM20,conf/cidr/Zaharia0XA21}, but also lack support for critical features~\cite{journals/pvldb/ArmbrustDPXZ0YM20,delta_issues}.

\labeltitle{Missing Features in OTFs.} This paper addresses the problem of how to support 
multi-query and multi-table 
transactions for LHs while providing strong ACID guarantees prominently discussed in the community \cite{delta4_multi_table,ducklake}. More specifically, we provide native support for the following three missing features: (i) \emph{Recovery}, (ii) \emph{Complex Transactions}, and (iii) \emph{Isolation}.
First, the conflict resolution of OTFs cannot track concurrent changes, often resulting in multiple aborts and higher transaction costs~\cite{delta_issues}. We refer to this feature as (i) \emph{Recovery}. Second, OTFs today do not support cross-operation dependencies, thereby only allowing simple business logic of single query transactions~\cite{journals/pvldb/ArmbrustDPXZ0YM20,conf/sigmod/LevandoskiCDDEH24,iceberg_specs,delta_doc,conf/cidr/Zaharia0XA21}. We refer to it as (ii) \emph{Complex Transactions}. 
Third, OTFs treat tables separately, leading to inconsistencies and dirty reads under concurrency~\cite{journals/pvldb/ArmbrustDPXZ0YM20}, and we refer to it as (iii) \emph{Isolation}. 
Unfortunately, the lack of support for these features prohibits concurrent writes in a system and can not guarantee properties like referential integrity across tables for concurrent reads. As a result, one cannot execute multi-query and multi-table transactions, forcing use cases for elastic compute or similar \cite{conf/icde/SethiTSPXSYJHSB19,spark_web,trino_web,conf/hotcloud/ZahariaCFSS10} to rely on synchronization methods that block the whole LH \cite{journals/pvldb/ArmbrustDPXZ0YM20,conf/cidr/Zaharia0XA21}. 
For example, concurrent analytics could observe a reference to a bank account that does not exist yet (see our running example in Section \ref{sec:metadata_design}).

\labeltitle{State of the Art.} 
Multi-query and multi-table transaction support is a highly requested feature for LHs \cite{journals/pvldb/ArmbrustDPXZ0YM20,delta4_multi_table,ducklake,conf/cidr/Zaharia0XA21} that recently got attention from DuckDB~\cite{ducklake} and Databricks \cite{delta4_multi_table}. In Table \ref{fig:overview}, we summarize the most notable approaches for multi-query and multi-table transaction support in LHs and their specific design decisions. In summary, most systems do not adapt to concurrent changes (cf. (i); Recovery)~\cite{conf/cidr/0001KPDSZ23,spark_doc,ib_reliability_docs,delta_doc}. At the same time, solutions for cross-operation dependency detection (ii) and global transaction isolation (iii) are either not present or currently in development for some systems.
Despite known techniques like the SAGA pattern \cite{conf/sigmod/Garcia-MolinaS87} or MVCC \cite{journals/tods/BernsteinG83,journals/pvldb/BernsteinRWY11,conf/ipps/IwabuchiSPEGM16,journals/pvldb/LiuHMCLSSSAA20,conf/sigmod/0001MK15} that ensure similar features in traditional database systems, all previous approaches listed in Table \ref{fig:overview}, add auxiliary components external to the object store but specific to a single OTF format~\cite{project_nessie_web,xtable_web,delta4_multi_table,ducklake} (D1). 
At the time of writing, only DuckLake~\cite{ducklake} and Delta Lake 4.0~\cite{delta4_multi_table} support multi-query and multi-table transactions in OTFs, but they diverge significantly from OTF principles by offloading all metadata to DuckDB~\cite {ducklake} and/or heavily modifying the OTF structure~\cite{delta4_multi_table}. Additionally, they are facing technical limitations, making their position in Table \ref{fig:overview} unclear (see Section \ref{sec:related_work}). However, relying on these external services raises concerns for interoperability across vendors, potentially reenabling vendor-lock in, contradicting the original idea of LHs as an open data warehouse \cite{journals/jpdc/ErramiHKB23,conf/cidr/Zaharia0XA21}.
Hence, we aim to support multi-table and multi-query transactions without any external structures required.

\labeltitle{LakeVilla (LV)} is our toolbox that implements the three features (i) recovery, (ii) complex transactions, and (iii) isolation to provide multi-table and multi-query transaction support under concurrency. LV defines a modular set of modules that are non-invasive and compatible with existing OTFs (D2). In contrast to prior work~\cite{project_nessie_web,xtable_web,journals/pvldb/ArmbrustDPXZ0YM20,ducklake}, LV's modules \#1--\#4.2 are integrated into the original OTF protocol and enable each transaction to choose its desired guarantees individually (D1). 
We ensure LV's correctness by guaranteeing that modules result in a valid state of the raw OTF. This allows a simple integration into existing LH systems and compatibility with clients running the raw OTF. More specifically, LV composes \#1--\#4.2 into three modular features:
First, LV[Recovery] (LV[R]) uses logical snapshot version markers (\#1) with physical and logical transaction sublogs (\#2) defining a greedy transaction model to support recovery (i).
Second, LV[Complex Transactions] (LV[CT]) introduces a marker-based conflict and deadlock detection model (\#3) among all tables to support complex transactions with an arbitrary number of queries/operations (ii).
Lastly, LV[Isolation] (LV[I]) introduces a global version log (\#4.1 and \#4.2) that checks global relations among tables and acts as a new layer on the object store above all tables (iii). 
We also demonstrate how each LV feature strengthens the ACID guarantees of LHs and discuss its implications.

\labeltitle{Key insights.} We show that LV guarantees serializability and linearizability for OTFs on the entire LH, not just individual tables or operations. The non-invasive integration of LV[I] (reads)
and 
LV[R, CT] (writes)
can be done with minimal overhead of 2.5\% (TPC-DS reads) and 2\% (YCSB writes) and incur negligible and predictable costs for typical LH workloads, like analytics.

\labeltitle{Our contributions:} In this paper, we give an overview of OTFs and their design area in Sections \ref{sec:metadata_design} and \ref{sec:background},

\begin{itemize}
    \item[C1] Define modules \#1--\#4.2 and compose them into the features LV[R], LV[CT], and LV[I] for (i)--(iii) providing compatibility across all their combinations and the base OTF without introducing vendor lock-in (Section \ref{sec:LakeVilla-Prototype}),
    \item[C2] Show that LV improves the ACID guarantees of OTFs involving multiple tables, queries, and concurrent writes, especially guaranteeing Snapshot Isolation up to Serializability (Section \ref{sec:Guarantees}),
    \item[C3] Introduce an LV prototype with minimal overhead and predictable costs when executing our adapted YCSB-LH~\cite{conf/cloud/CooperSTRS10}, CAB-LH~\cite{journals/pvldb/RenenL23}, TPC-C~\cite{tpcc}, LSTBench~\cite{2024lstbench}, and microbenchmarks (Section \ref{sec:evaluation}).
\end{itemize}
Afterward, we discuss related work and conclude in Sections \ref{sec:related_work} and \ref{sec:conclusion}.

\section{Lakehouses \& Open Table Formats}
\label{sec:metadata_design}

\labeltitle{Lakehouses} (LHs) are an evolution from Data Lakes~\cite{journals/pvldb/ArmbrustDPXZ0YM20,conf/cidr/0001KPDSZ23,conf/cidr/Zaharia0XA21} for efficient processing of structured and unstructured data in object stores using a query engine~\cite{conf/sigmod/BehmPAACDGHJKLL22,spark_web,trino_web}.
LHs aim for comparably lower storage costs in data lakes while maintaining reasonable performance at the expense of data freshness for their expected workloads.
The typical use cases of LHs are machine learning and analytics with concurrent writes from various data sources with unique characteristics~\cite{journals/pvldb/ArmbrustDPXZ0YM20,Iceberg_website,delta_lake_web,conf/cidr/Zaharia0XA21}. This work focuses on analytical applications with concurrent writes, typically consisting of an initial load phase and mixed workloads of 15\% continuous writes, 30\% analytical reads, and 55\% read-writes as observed by industry and publicly available datasets (\eg Snowset~\cite{snowflake-nsdi20}, Redset~\cite{Renen2024}). Writes are typically split between 80\% inserts and 20\% updates~\cite{DBLP:conf/sigmod/SchmidtKHSK24}. 
Across SAP use cases from Ariba~\cite{sap_ariba}, S4/HANA~\cite{sap_hana}, and SAP Sales Cloud~\cite{sap_sales_cloud}, we found initial load sizes of multiple GBs up to single TBs.
Workloads have ingestion rates of x10\textsuperscript{3} kB to MB-sized events per hour and complex analytics with up to x10\textsuperscript{2} to x10\textsuperscript{3} concurrent users.
The required data freshness lies between a few seconds to daily.

\labeltitle{Open table formats (OTFs)} on the object store are the basis of LHs \cite{hudi_web,Iceberg_website,delta_lake_web}. Those formats add a separate metadata layer for each table, organize data, and control client concurrency. This enables advanced analytical features, easy adaptation, and builds on open-source formats \cite{journals/pvldb/ArmbrustDPXZ0YM20,conf/cidr/0001KPDSZ23,hudi_web,Iceberg_website,delta_lake_web,conf/cidr/Zaharia0XA21}. Typically, OTFs use a file hierarchy and rely on atomic requests to the object store \cite{journals/pvldb/ArmbrustDPXZ0YM20,conf/bigdataconf/BegoliGK21,conf/cbd/ChenSLLJ22,conf/cidr/0001KPDSZ23,conf/cidr/Zaharia0XA21}. Two well-known OTFs \cite{conf/cidr/Zaharia0XA21} (\ie Snowflake \cite{snowflake_iceberg_web}, Amazon EMR~\cite{lakehouse_aws_emr}, BigQuery~\cite{delta_lake_bigQuery,conf/sigmod/LevandoskiCDDEH24}) are Delta Lake and Iceberg \cite{Iceberg_website,delta_lake_web}.

The metadata layers of OTFs are essential and the biggest difference from previous architectures \cite{conf/cidr/Zaharia0XA21}. LV operates directly within the OTFs used by LHs. For simplicity, this paper uses Delta Lake to explain LV. However, LV can be analogously applied to Apache Iceberg \cite{Iceberg_website} or other OTFs with minimal changes.

\begin{figure}
    \centering
    \includegraphics[width=0.95\linewidth]{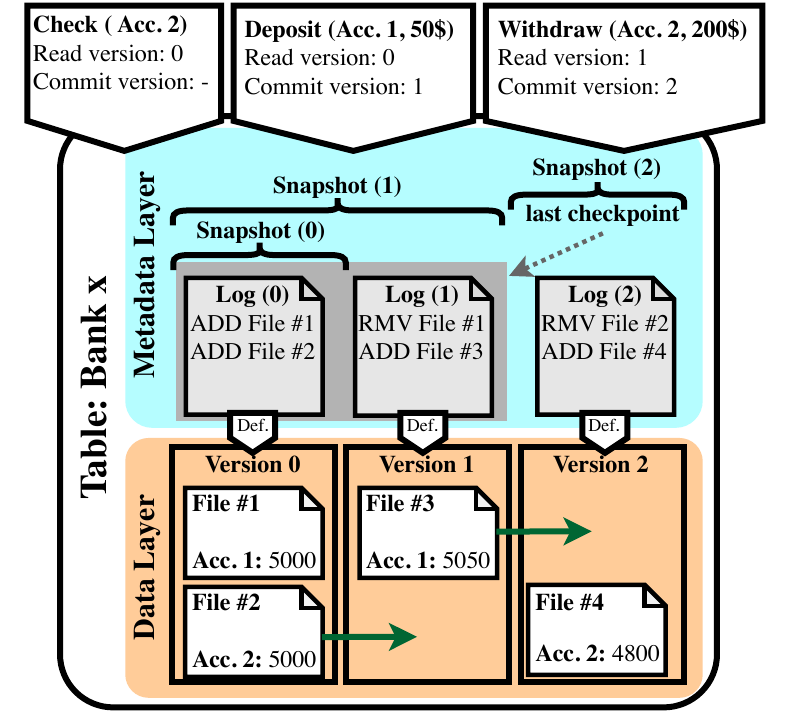}
    \caption{Metadata Layer design of Delta Lake in a banking example (running example) (draw.io)}
    \label{fig:dl-design}
\end{figure}

\labeltitle{Delta Lake \cite{delta_lake_web}} is one of the most well-known and practically used OTFs \cite{conf/cidr/Zaharia0XA21} (\ie Amazon EMR~\cite{lakehouse_aws_emr}, BigQuery~\cite{delta_lake_bigQuery,conf/sigmod/LevandoskiCDDEH24}). 
The structure of a delta lake table \cite{delta_lake_web,delta_doc} is depicted in Figure \ref{fig:dl-design} for a fictional table Bank x. A Delta Lake table consists of a metadata layer structured like a linked list and a data layer. A snapshot is defined by its respective log file and encodes all file operations to advance the previous snapshot to its own \cite{journals/pvldb/ArmbrustDPXZ0YM20}. These operations range from adding or removing files to adding metadata, and more \cite{journals/pvldb/ArmbrustDPXZ0YM20}. To retrieve the full table version of a snapshot, engines must access the respective logs. Figure \ref{fig:dl-design} defines snapshots 0, 1, and 2 with logs (0), (1), and (2). Table version 0 is fully defined by Log (0) and contains Files \#1 and \#2, while version 1 is defined by Logs (0) and (1) and consists of Files \#2 and \#3. 
To avoid excessive sequential accesses to all logs, Delta Lake implements a pointer to the last checkpoint created, usually stored in a separate file, and regular checkpoints stored in Parquet format \cite{journals/pvldb/ArmbrustDPXZ0YM20}. A checkpoint summarizes all valid table instructions and snapshots at the time of its creation \cite{journals/pvldb/ArmbrustDPXZ0YM20}. In Figure \ref{fig:dl-design}, the last checkpoint contains the instructions of logs (0) and (1). Hence, a client only reads the checkpoint and Log (2) to retrieve version 2.

\subsection{Running Example: Delta Lake Operations}

\labeltitle{Reads: Check.}
We showcase the operations within a Delta Lake table using a simple banking example shown in Figure \ref{fig:dl-design}. All accounts of bank x are stored within the table and contain their id and balance. First, the bank retrieves a check request for account 2. Hence, the used engine retrieves the current snapshot by issuing a LIST request on the metadata directory and determines the required logs \cite{journals/pvldb/ArmbrustDPXZ0YM20}. For snapshot (0), it only reads Log (0) and accesses Files \#1 and \#2. After the required data is retrieved, the engine can commit without changing the metadata layer.

\labeltitle{Writes: Deposit and Withdraw.}
The first write request the bank retrieves is a deposit request for account 1 of 50\$. The engine starts exactly like a read request by determining the newest snapshot, reading the respective logs, and assembling the current version. However, since files can only be overwritten in data lakes, it continues to create a new file \#3 with the updated balance of 5050\$ for account \#1. When the operation is committed, the engine adds log (1), removing file \#1 with the outdated value and adding \#3. This effectively creates snapshot (1). Due to the engine settings, it creates a checkpoint containing the information of all current logs (0 and 1) and sets the last\_checkpoint pointer.

Next, 200\$ is withdrawn from Acc. 2. Similar to before, an engine receives snapshot (1) but reads the last checkpoint instead of Log (0) and (1). The required file \#2 is used as the basis for the updated balance 4800\$ written as file \#4. During commit, the engine creates snapshot (2) by adding Log (2), replacing File \#2 with \#4. Here, the engine decides not to create a checkpoint and finishes its operation.

\section{The Lakehouse Design Area}
\label{sec:background}

\subsection{Missing Features of OTFs}
\label{subsec:missing-features}

Despite different metadata layers, all OTFs share similar missing features by default, as also acknowledged in prior work~\cite{journals/pvldb/ArmbrustDPXZ0YM20,delta4_multi_table,conf/cidr/Zaharia0XA21}. In this paper, we cover:

\labeltitle{(i) Recovery: No Duplicated Work on Aborts.}
All OTFs define a single table-level synchronization point in their metadata layers in the form of appending logs or branches to enable optimistic concurrency control~\cite{journals/pvldb/ArmbrustDPXZ0YM20,iceberg_specs}. Hence, a transaction can only be committed if it successfully passes this bottleneck. 
All transactions based on the same accessed branch/log must abort as the new commit might have accessed the same data \cite{iceberg_specs}. 
However, OTFs do not provide any means to recover or undo a failed transaction. This can result in unreferenced data files, additional costs, and high latency. Redoing the transaction results in redundant work in the form of additional requests to the object store, especially uploading duplicate data files. The impact of this missing feature can intensify for more complex operations, as they might starve or fail with the standard OTF concurrency control, as seen in section \ref{sec:evaluation} for TPC-C.

\labeltitle{(ii) Complex Transactions: Concurrent Multi--query and Multi--table Transactions.} 
The OTF design decision to focus on transactions consisting of a single SQL query (single-query) and accessing strictly one table (single-table) led to publishing each (sub-) operation after its execution (a) and a table-level design (b) of the metadata~\cite{journals/pvldb/ArmbrustDPXZ0YM20,iceberg_specs,conf/cidr/0001KPDSZ23,delta_doc}. 
(a) breaks the ACID guarantees for multi-query transactions:
When simply applying the current designs, intermediate changes of multi-query transactions would become immediately visible (violating atomicity) and could get overwritten by other concurrent transactions (violating consistency and isolation). 
Further, (b) prevents multi-table transactions:
An abort in one table could lead to a partially committed transaction (violating consistency) that can not be undone without potentially aborting newer transactions or deleting committed values (violating durability and introducing a risk of cascading aborts). 
Thus, users must either break isolation or risk starvation of complex queries for complex transactions.

\labeltitle{(iii) Isolation: Controlling global versions.} 
Due to the independent metadata layers in OTFs, multi-table transactions must interact with the metadata layer of each table separately \cite{Iceberg_website,iceberg_specs,delta_lake_web,delta_doc}. This design decision ignores any global information about the tables' states and dependencies, assigning versions on the table level.
Under concurrency, other transactions might observe an inconsistent combination of table versions (dirty read), as some tables might adopt changes of another transaction earlier than others. Most prominently, LHs cannot prevent read transactions from observing potentially partially committed multi-table transactions even with the restriction to a single writer \cite{conf/cidr/Zaharia0XA21}.
A global lock, as suggested by \cite{journals/pvldb/ArmbrustDPXZ0YM20,iceberg_specs,delta_doc}, can solve the problem but reduces the LH's throughput due to partially enforcing sequential execution.
As concurrent reads and read-writes are the dominant workload of LHs, it becomes essential to guarantee transaction isolation of read and write transactions.

\subsection{Design Questions}

When addressing missing LH features and exploring prior approaches (e.g., \cite{delta4_multi_table,ducklake}), we observed four design directions described as questions:

\labeltitle{(Q1) What concurrency mechanism should we use?} 
Exclusive locking~\cite{journals/pvldb/ArmbrustDPXZ0YM20} prevents concurrent writes while aborting all conflicts. A branching solution, like Project Nessie~\cite{project_nessie_web}, isolates transactions in different branches  \cite{iceberg_nessie_doc}.
Lastly, Delta Lake's newly announced commit owner \cite{delta4_multi_table} and DuckLake \cite{ducklake} apply \textit{versioning-based} conflict detection and resolution \cite{journals/pvldb/ArmbrustDPXZ0YM20,Iceberg_website} known from DB research \cite{journals/tods/BernsteinG83,books/aw/BernsteinHG87,conf/sigmod/0001MK15}. 

\labeltitle{(Q2) Do we implement an internal or external solution?}
Most LH projects extract responsibilities from the OTF to other external services above the object store \cite{delta4_multi_table,project_nessie_web,ducklake,xtable_web} or to their database \cite{ducklake}. This essentially provides a simple and performant solution.
In contrast, a \textit{built-in} solution within the OTF design without auxiliary architectures reduces dependencies on such services.

\labeltitle{(Q3) Should we work with existing OTFs or redefine them?}
Delta Lake 4.0 updates its original design and provides backward compatibility with its backfilling feature \cite{delta4_multi_table}. Further, DuckLake manages metadata exclusively within their database, such that only the raw data files remain on the object store \cite{ducklake}. In contrast, \textit{adaptive} approaches like XTable \cite{xtable_web} and Project Nessie \cite{project_nessie_web} integrate into existing OTF architectures without noticeable structure changes, providing easier adoption on already deployed systems.

\labeltitle{(Q4) For which type of transactions do we use which solution?}
Most LH projects are one-fit-all solutions affecting all transaction types \cite{journals/pvldb/ArmbrustDPXZ0YM20,delta4_multi_table,project_nessie_web,ducklake,xtable_web}. 
This contradicts a \textit{modular} solution, providing individual choice between guarantees.
However, this limits its simplicity, as such solutions must offer modular building blocks. 

\subsection{Previous Approaches}

Regarding the missing features we identified in Subsection \ref{subsec:missing-features}, we found four relevant designs: External locks~\cite{journals/pvldb/ArmbrustDPXZ0YM20}, Project Nessie \cite{project_nessie_web}, Delta Lake 4.0 commit owners~\cite{delta4_multi_table}, and DuckLake~\cite{ducklake}, each representing current approaches.

\labeltitle{External Locks \cite{journals/pvldb/ArmbrustDPXZ0YM20}} follow a locking mechanism (Q1) to sequentialize LH transactions on a table. It is established via an external service like Zookeeper (Q2) and works with the existing OTF structure (Q3). However, due to its table-level granularity, it provides a solution that must be pushed to all (write) transactions (Q4). However, while solving the missing feature (iii), it completely removes concurrency.

\labeltitle{Project Nessie \cite{project_nessie_web}} follows similar decisions for Q2 -- Q4, but chose a branching mechanism (Q1). It establishes git-like logic for Iceberg, but currently does not allow merging different branches. However, this is essential for the missing feature (ii).

\labeltitle{The Delta Lake Commit Owner \cite{delta4_multi_table}} combines versioning and branching (Q1) to provide a central commit via an external, proxy-like, service (Q2). In addition to their external service, they adjust the OTF to fit their design tightly (Q3). While it effectively improves Delta Lake, it also causes compatibility issues with existing implementations. The announced backfilling feature tries to compensate for that, but causes more complexity and delays~\cite{delta4_multi_table}. Lastly, as this approach centralizes commits, the service must be used for all transactions (Q4).

\labeltitle{DuckLake \cite{ducklake}} implements multi-table and multi-query transactions by separating the metadata layer from the object store. It manages the metadata of the LH within DuckDB to provide a versioning-based (Q1) solution externally from the object store (Q2). Thus, DuckLake splits the LH design into two distinct parts (Q3), which forces all transactions to go through their structure in DuckDB (Q4).
However, we believe such solutions go against the original concept of LHs as they rely on external services, hide the object store behind an external bottleneck, and might reintroduce vendor lock-in.

\section{LakeVilla}
\begin{figure}
    \centering
    \includegraphics[width=\linewidth]{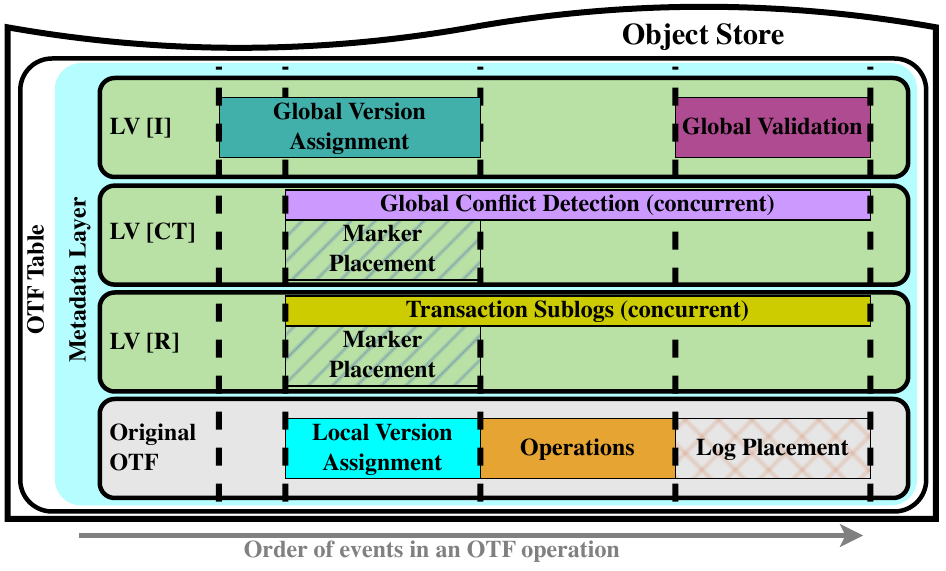}
    \caption{Overview of all LakeVilla (LV) features and how they integrate into the original OTF (draw.io)}
    \label{fig:lv-overview}
\end{figure}

\label{sec:LakeVilla-Prototype}
\labeltitle{LakeVilla (LV)} is our \emph{versioning-based} (Q1), \emph{built-in} (Q2), \emph{adaptive} (Q3), and \emph{modular} (Q4) OTF extension providing the missing features (i) -- (iii). 
LV defines three feature sets: LV[R] for \emph{(i) Recovery}, LV[CT] for \emph{(ii) complex transactions}, and LV[I] for \emph{(iii) isolation}. 
Our design can seamlessly integrate into OTFs such as Apache Iceberg and Delta Lake without reformatting or reconfiguring the whole system.

\labeltitle{The Architecture of LV.} LV defines modules 
compatible with OTF operations. Its design does not impact the baseline structure to allow concurrent access to the LH with legacy readers or even other LV combinations. Figure \ref{fig:lv-overview} summarizes the typical OTF structure and which modules each LV feature adds. LV[R] and LV[CT] add 'marker placement' to extend the 'Local version assignment' of the OTF. Further, LV[R] incorporates a transaction sublog system, and LV[CT] defines concurrent conflict detection across tables. Lastly, LV[I] implements a global layer that assigns ('Global Version Assignment') and validates versions and changes ('Global Validation') to guarantee isolation. This section describes all LV features and their modules.

\subsection{LakeVilla[Recovery] (LV[R])}
\label{subsec:LV-R}

\labeltitle{Motivation: Why OTF transactions abort.} 
When operating with concurrent transactions, the key challenge to prevent aborts is adapting to changes in the table versions of the metadata layer. By default, OTFs apply optimistic concurrency control that checks for conflicts at the commit of a transaction \cite{ib_reliability_docs,delta_doc,conf/cidr/Zaharia0XA21}. This approach assumes that most operations do not conflict. However, OTFs also provide table-level granularity, meaning that all concurrent write operations on the same table will conflict without fine-grained conflict management. Hence, we propose LV[R] to provide recovery (i). This feature combines two modules we designed for OTFs following our decisions for Q1--Q4: Our marker placement (module \#1) reserves snapshot IDs at a transaction's beginning seamlessly and publishes a transaction's changes atomically. To adapt to concurrent changes and faults, we propose our transaction sublogs (module \#2).

\begin{figure}
    \centering
    \includegraphics[width=0.45\textwidth]{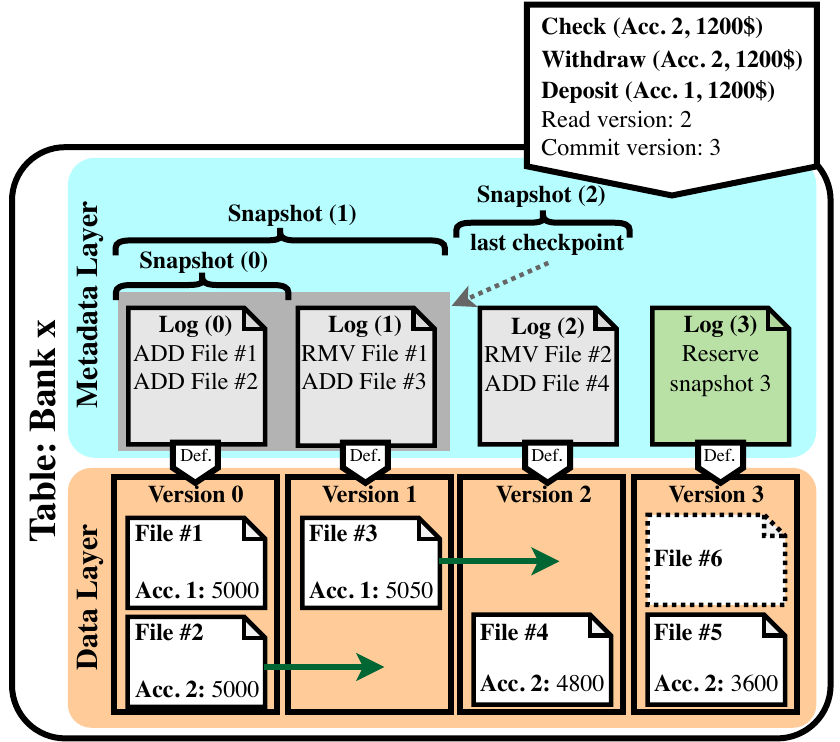}
    \caption{Module \#1: Markers (green) reserve snapshot versions in the metadata layers (draw.io)}
    \label{fig:technique1}
\end{figure}

\labeltitle{Module \#1: Marker Placement.}
We propose file-based markers within OTF metadata layers to prevent concurrent OTF transactions from claiming snapshot versions.
We ensure their uniqueness and atomicity by using Put-If-Absent or conditional writes to the object store already used in OTFs \cite{aws_conditional_writes,journals/pvldb/ArmbrustDPXZ0YM20}. Additionally, we implement continuous checks for object stores without atomic requests, accessing each marker's HEAD to detect potential overwrites. If a transaction fails to claim a version or its marker was deleted by someone else, it aborts or claims a new marker. Markers are temporary metadata objects that must be deleted or overwritten.

\labeltitle{Markers in OTFs.}
In a Delta Lake table, a marker is appended to the log directory. In our example table Bank x, shown in Figure \ref{fig:technique1}, reserves log (3) and its snapshot (3). The executing transaction, consisting of check, withdraw, and deposit operations, can now fully execute and publish all its changes into Log (3) without snapshot (3) being claimed by anybody else. This effectively implements a write lock~\cite{books/aw/BernsteinHG87,journals/pvldb/BernsteinRWY11,conf/ipps/IwabuchiSPEGM16,journals/pvldb/LiuHMCLSSSAA20} into OTFs.  
As most OTFs define a synchronization point in their metadata layers \cite{journals/pvldb/ArmbrustDPXZ0YM20,hudi_web,iceberg_specs}, their marker integration is analogous to the one presented here: For example, in Iceberg, we place markers in the area of manifest lists and link to the snapshot.

\labeltitle{Marker Compatibility with Other Clients.}
Following our design principle of \textit{modularity} (Q4), we propose that markers not enforce the same guarantees or new techniques on all transactions. Hence, our markers must fulfill their purpose while still being compatible with the original OTF protocols. LV's markers use the extensible open-source formats (\eg JSON) of the original logs/manifest and appear empty for the original protocols. At the same time, they are detectable for our marker-based clients with the bookkeeping data contained. This data consists of the transaction ID that created the marker and additional information required for other LV modules (\eg dependent transactions, time of creation, ...).

\labeltitle{Marker Commit.}
Nevertheless, commits using markers should be atomically visible to all clients. So, we designed a commit phase that replaces these markers noticeable to legacy readers: LV collects and summarizes all relevant information according to the OTF's required instructions during a commit (see Delta Lake in section \ref{sec:metadata_design}). If needed, it builds underlying structures, such as Iceberg's manifest lists, and finally overwrites the placed marker with all the necessary instructions. This action effectively makes the whole transaction result of a table available even for the original OTF protocols. In case of an abort, our client overwrites a marker with a previous manifest list or an empty log entry to not disturb the original protocols.

\begin{figure}
    \centering
    \begin{subfigure}{.235\textwidth}
        \centering
        \includegraphics[width=0.95\textwidth]{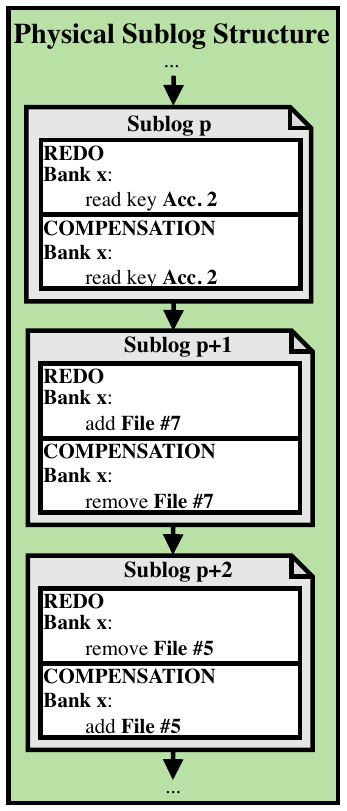}
        \caption{Physical }
        \label{fig:physicalSAGA}
    \end{subfigure}%
    \begin{subfigure}{.235\textwidth}
        \centering
        \includegraphics[width=0.95\textwidth]{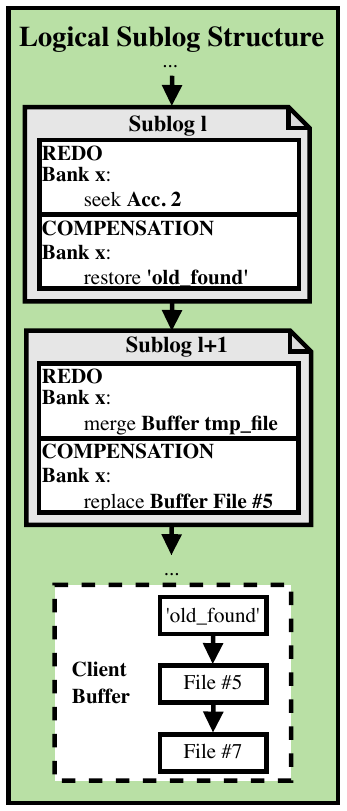}
        \caption{Logical}
        \label{fig:logicalSAGA}
    \end{subfigure}
    \caption{Module \#2: Logical and physical Sublog instructions for "UPDATE \textbf{Bank x} SET \textbf{Balance=100} WHERE key=\textbf{Acc. 2}" (draw.io)}
    \label{fig:SAGA}
\end{figure}

\labeltitle{Module \#2: Transaction Sublogs.}
After introducing our marker placement, we propose a transaction recovery system using our sublogs. This module generates a structure of partial logs for running transactions, enabling Undos and Redos of operations. We designed our sublog structure and its sublogs from scratch because the underlying operations differ in file-based OTFs from database systems:
Operations generate, swap out, and split files to create a complete version for the following snapshot. 
Delta Lake's time-traveling \cite{journals/pvldb/ArmbrustDPXZ0YM20} already defines basic redo information for OTFs in its logs. However, its definition is static, losing information about the changes made, while a transaction must adapt in a conflict. For example, if a transaction adds 50\$ to account 1, the Delta Lake logs would only contain information about the new file created (see section \ref{sec:metadata_design}). However, this transaction must do a complete rollback if a concurrent operation is faster. Hence, we need a new logging type for redo operations in a concurrent LH environment. LV defines \emph{physical} and \emph{logical} sublogs to provide the necessary information for this module. Figure \ref{fig:SAGA} shows the generated LV sublogs in our \emph{physical} and \emph{logical} definition for the query "UPDATE Bank x SET Balance=100 WHERE key=Acc. 2" executing withdraw(Acc. 2, 3500\$) in our running example.

\labeltitle{LV's Physical Sublogs} describe the explicit changes made to the previous OTF table state, such as adding and removing files.
We show an example of a physical sublog for the above query in Figure \ref{fig:physicalSAGA}.
The first instruction in Figure \ref{fig:physicalSAGA} reads the file with the specified key 'Acc. 2' for the selected table 'Bank x', resulting in the physical path of the found data file 'File \#5' and the file's contents. The respective compensation step repeats this operation, as the redo operation does not change the overall table state. For the next step, the client uploads the new data file under the newly generated path 'File \#7' containing 'Balance=100' and defines its removal for the compensation step. Lastly, the transaction removes the old data file at 'File \#5' and adds it for compensation.

\labeltitle{LV's Logical Sublogs} encode the example update query as shown in Figure \ref{fig:logicalSAGA}.
Its first sublog defines the search of the current file, which specifies the balance of 'Acc. 2' and stores the result 'File \#5' in the client's buffer. The respective compensation step logs the old value of the buffer 'old\_found' to restore the old buffer state. The second sublog references the current buffer value from step l ('File \#5') and merges the referenced file with the generated file 'tmp\_file' containing an instruction ('set') and modifier ('balance=100') to generate the updated version. The path of the resulting file is stored in the client's buffer. For compensation, we replace the merged file with the buffer value (File \#5). A transaction must ensure that it executes a redo on all its logical sublogs before it commits.

\labeltitle{The File-based Redos and Undos of \#2.} The defined sublog structure allows a transaction to jump to any point in its execution by performing undos and redos: An undo traverses the sublog structure from the newest to the oldest sublog, executing the compensation steps to restore previous states. For our example in Figure \ref{fig:SAGA}, it effectively restores the transaction state before the update in both logging versions: For the \emph{physical} instructions, it adds File \#5 and removes the updated file File \#7 (compensation steps p+1 and p+2). Furthermore, as a read operation did not change the physical state of the table, we can skip the respective compensation (compensation step p).

The \emph{logical} version starts by replacing the path stored in its buffer with File \#5 and pushing 'File \#5' to the buffer (compensations step l+1). Internally, we implement replacements through the removal of File \#7 and the addition of File \#5 (similar to \emph{physical} instructions). Lastly, we must store the old buffer value 'old\_key' in the client's buffer for the following undo instructions (compensations step l). In contrast to undo, redo operations traverse the sublog structure from the oldest to the newest file, executing the redo steps. This process effectively results in a client executing the same operations performed initially on the current table version.

\labeltitle{LV[R]: A Combination of \#1 and \#2.}
The goal of our sublog module (\#2) is to recover from any failure and, in combination with module \#1, be able to remove markers of crashed clients. Further, LV[R] switches between \emph{physical} and \emph{logical} sublogs to manage concurrent writes and provide less overhead depending on the marker placement:
By default, LV[R] uses physical sublogs. As physical sublogs describe file operations that have already been performed on the object store, the respective operation must be assigned its intended OTF snapshot version, requiring no further computation. Therefore, a transaction using physical sublogs has precedence over logical sublogs. LV[R] reserves physical sublogs for operations without preceding concurrent transactions. If a transaction detects another preceding running transaction accessing the same table by its marker, it must switch to logical sublogs. This choice gives us two advantages: a transaction must not freeze until all preceding transactions are finished, and the system can invisibly execute operations that can not conflict with other transactions (e.g., blind appends). Hence, a transaction generates the file and adds a physical sublog for such cases, even in a logical structure. However, a transaction can never fully switch back from logical to physical logs. This mixture of modules \#1 and \#2 enables higher throughput in the metadata layer and transforms our markers from resembling exclusive write to shared write locks.

\labeltitle{LV[R]'s Conflict resolution and file-granularity.} To determine whether to commit or abort, LV[R] uses the markers placed in the metadata layer of a table. The unique and distinct order among the atomic markers ensures an ordering among all operations within a table. For concurrent transactions, the conflict detection triggers as soon as a preceding transaction replaces its marker. The waiting client accesses the content of the new log and checks if the prior transaction changed the same files or keys using its logical file-based sublog. LV[R] solves such file or key conflicts by undoing the transaction to the first operation where a conflict was detected and triggering a redo on the new table state. However, logical conflicts like dependencies across transactions or tables might trigger more prolonged recovery procedures, potentially completely re-executing the transaction. To alleviate this, we added a retry counter that interrupts recovery and aborts the transaction if an error appears multiple times. As an additional benefit of \#1 and \#2, their combination operates on file references, giving us file-based conflict granularity for all transactions using LV[R].

\labeltitle{LV[R]'s Client failure recovery.} 
Besides LV[R]'s conflict resolution during commit, it guarantees recovery from other causes of failure. Some common issues for OTFs revolve around upload/download failure, client crashes, or similar problems. Such failures might leave the LH metadata layer in an inconsistent or invalid state. Further, adding or removing files for updates or using markers and other LV modules amplifies this problem. LV[R]'s sublogs are also a safety mechanism that allows clients to undo all failed transaction changes even if they did not initiate them. To guarantee recovery in such situations, LV[R] requires that sublogs are only removed after clients replace their marker with a snapshot/log. Currently, LV uses timeouts and counters for failed transactions to detect crashed clients with active markers. If a client decides to free the markers of a potentially crashed client, it does not delete the sublog. This rule enables the crashed client to automatically claim a new marker and continue the transaction. 

\subsection{LakeVilla[Complex Transactions] (LV[CT])}

\labeltitle{Motivation: Why OTFs only support transactions on a single table.} As described in Section \ref{sec:metadata_design}, OTFs define distinct metadata layers for each table. Hence, all operations on different tables trigger independent operations in different metadata layers. This effectively blocks any consistency guarantees while writing to multiple tables: In our running example, we define the transfer operation across distinct banks (represented as different tables). This operation comprises our standard operations: Check, Withdraw, and Deposit. Across different tables, Withdraw and Deposit would interact with completely different metadata layers without any guarantee that both can commit or are ordered in the same way for concurrent transactions globally. Hence, OTFs do not provide any means to track such dependencies, which can cause inconsistencies and atomicity violations.

\begin{figure}
    \centering
    \includegraphics[width=0.45\textwidth]{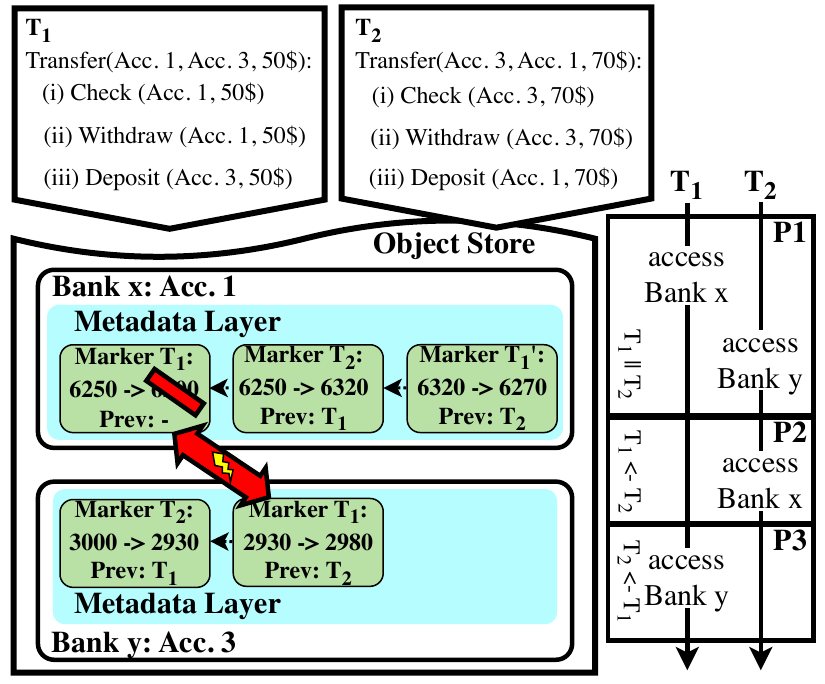}
    \caption{Module \#3: A marker-based conflict and deadlock detection model. This figure shows a conflict between tables 1 and 2 and how T\textsubscript{1} recovers from it (draw.io)}
    \label{fig:technique3}
\end{figure}

\labeltitle{Why can we not reuse LV[R]?} 
Simply re-applying LV[R]'s marker idea only partially solves the above problems, as multi-table transactions can still break the ordering of transactions operating on the same bank(s) (=related transactions) as implied by the switched ordering of markers in Figure \ref{fig:technique3}. The metadata layer of one table would not be able to recognize any dependencies between tables, thereby enabling inconsistencies, deadlocks, and partial aborts. Hence, we designed LV[CT] that tracks the dependencies across tables by using markers (\#1) to expose the transaction globally and its new module 'global conflict detection' (\#3). This combination allows us to implement multi-table transactions that guarantee the same ordering of related transactions across all tables.

\labeltitle{Module \#3: Global Conflict Detection.} 
For multiple tables, we can encounter a deadlock or cause inconsistencies when placing more markers into these tables and waiting for preceding transactions (compare deadlocks for locks). Hence, module \#3 defines a global conflict and deadlock detection model for markers. Our goal is to track dependencies across tables without causing deadlocks among transactions. Our strategy builds a client-based directed acyclic graph (DAG) structure of dependencies, described in four simple rules:

\begin{itemize}
    \item[(I)] Always use the most recent public version of a table
    \item[(II)] Always expose your preceding transactions (outgoing DAG edges) within all of your markers
    \item[(III)] Check your following markers if they reference yourself (incoming DAG edges).
    \item[(IV)] Free your marker if you detect a cycle or your timeout runs out.
\end{itemize}

In summary, this module defines a strategy to check transactional dependencies concurrently. It requires transactions to acquire markers in every table accessed (I). Those markers must contain information about the respective transaction's dependencies (II) and concurrently check for changes in detected markers of other transactions (III). If a cycle is formed when accessing a new table, the transaction must free its own markers and set new markers, inverting its outgoing DAG edges (IV). We call this procedure 'marker shift'.

\labeltitle{Example: Phase P1.}
We demonstrate our strategy with the example in Figure \ref{fig:technique3} for two transfer transactions: Transactions T\textsubscript{1} and T\textsubscript{2} check the balance of their sender's account. Since Acc. 1 and Acc. 3 are registered at different banks, they start by accessing different tables and placing their markers in the respective metadata layer. These transactions can be considered unrelated at this point.

\labeltitle{Example: Phase P2.}
After withdrawing 70\$ from Acc. 3, T\textsubscript{2} accesses bank x to deposit the amount to Acc. 1 and places its marker behind T\textsubscript{1}'s marker in bank x. When initializing the new marker and reading the table state, T\textsubscript{2} recognizes T\textsubscript{1}'s marker. As this relation does not violate its observed dependencies, T\textsubscript{2} continues. It overwrites all its markers in Bank x and y with its dependency list \{T\textsubscript{1}\}. T\textsubscript{2} depends now on T\textsubscript{1}. 

\labeltitle{Example: Phase P3.}
Later, T\textsubscript{1} accesses bank y to deposit 50\$, places its marker in the metadata layer of Bank y, and checks for other markers. T\textsubscript{1} detects T\textsubscript{2}'s marker and retrieves its data. It compares its dependency list with T\textsubscript{2}'s list and checks for potential deadlocks and conflicts. Because T\textsubscript{2} lists T\textsubscript{1} as dependent and T\textsubscript{1} would depend on T\textsubscript{2} after the operation, it concludes that a deadlock might occur and triggers a marker shift. Hence, T\textsubscript{1} frees its old marker in bank x by overwriting it with an empty manifest/log and places a new marker "Marker T\textsubscript{1}'" into the respective metadata layer. After confirming that the dependencies match across tables and merging the changes by T\textsubscript{2}, T\textsubscript{1} continues its operation. If T\textsubscript{2} decides to access another table and finds a conflict contradicting its internal observations, it will reevaluate these observations. Here, it checks the origin of its dependency on T\textsubscript{1} (here: Bank x) if the situation changes. Depending on the result, it updates its internal observations or solves the conflict.

\labeltitle{LV[CT]: Combining \#1 and \#3}
In total, we define LV[CT] for transactions on multiple tables by implementing \#1 and extending it with \#3.  
Like before, LV holds a marker to lock a table until the entire transaction is processed and overwrites it on commit/abort with a real manifest/log to unlock (compare locks~\cite{books/aw/BernsteinHG87}). 
However, LV[CT] adds the condition that a marker is only overwritten during a marker shift or all its markers during commit.

\labeltitle{LV[CT]: Use-Case, Larger Cycles, and delay handling.} Module \#3's conflict detection effectively detects and resolves smaller dependency cycles. However, due to larger DAG cycles, delays of marker checks, marker generation, or general network delays, LV[CT] can not guarantee that all markers are up to date at all times. Hence, our file-based deadlock detection model must consider such uncertainties.
Nevertheless, we found the approach suitable for the LH use-case (Section \ref{sec:metadata_design}). Module \#3 can be executed concurrently to write transactions, does not disturb the original OTF structure, and has no impact on analytical queries (30\% of its queries). Additionally, current designs do not allow concurrent writes on the same table, which makes a marker shift an emergency method that is rarely executed. In case some of the above scenarios impact the design, like in unusual LH workloads (see TPC-C evaluation), we introduced a timeout to markers and waiting transactions to prevent larger cycles or unresolved markers in our implementation to avoid the overhead of building larger DAGs. If a timeout runs out, the respective transaction triggers a marker shift on all its tables, breaking incoming DAG edges to other transactions.

\subsection{LakeVilla[Isolation] (LV[I])}
\label{subsec:LV-I}

\labeltitle{Motivation: Why do we need a global LH Layer?} 
LV[CT] successfully manages transactional dependencies across tables by using markers (\#1). This idea works well for write-heavy operations but introduces unnecessary overhead when applied to read-only or read-write transactions: LV[CT] would force such transactions to manage markers and thus write empty logs during commit to the OTF metadata layer. This introduces unnecessary writes and latency to such transactions.
Hence, we introduce LV[I], defining a lightweight global layer that manages commits and global dependencies (modules \#4.1 and \#4.2) for such transactions. This change guarantees globally consistent versions at all times and global isolation (iii). Furthermore, it expands steps from the original OTFs based on MVCC, like version assignment and validation \cite{journals/tods/BernsteinG83}. Our approach minimizes cost and avoids slow requests like LIST to the object store. 

\begin{figure}
    \centering
    \includegraphics[width=0.45\textwidth]{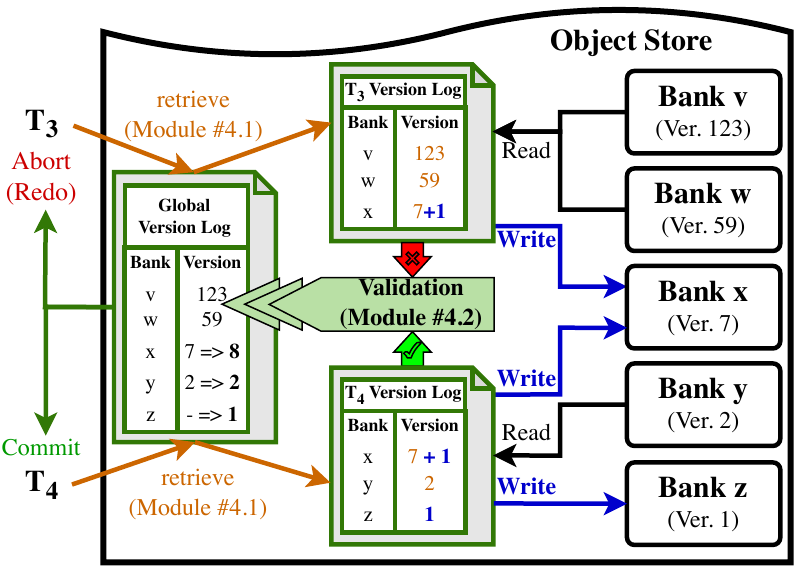}
    \caption{Modules \#4.1 and \#4.2: A global version log assigns snapshots and checks compatibility during commit. T\textsubscript{2} succeeds and updates Banks x, y, z (draw.io)}
    \label{fig:technique4}
\end{figure}

\labeltitle{Module \#4.1: Global Version Assignment} defines a global layer for object stores that manages all table snapshot versions to track and isolate transactions globally. Module \#4.1's global version log summarizes all tables' most recent valid versions and exposes this information in a single file at a static object store key.
If a client accesses a table, it is bound to the versions received by the global version log and can only use table versions equal to or smaller than the specified version. The example in Figure \ref{fig:technique4} shows a global version log that tracks different bank tables and their newest snapshot versions. T\textsubscript{3} and T\textsubscript{4} both start their execution and retrieve the version combination (Bank v, Ver. 123), (Bank w, Ver. 59), (Bank x, 7), and (Bank y, Ver. 2) from the global version log.

\labeltitle{Module \#4.2: Global Validation.}  
To check the transaction's compatibility with the current global version during its commit, we propose a file-based validation phase similar to MVCC's check \cite{journals/tods/BernsteinG83,conf/sigmod/0001MK15}. A validation of a transaction starts by checking and atomically copying its version changes to a shared, temporary key on the object store (PUT-IF-ABSENT). If the transaction succeeds, it iterates over its internal transaction version log and compares the version numbers of the current global version log. 
However, table versions in LHs are different. We can distinguish between a table's read version (snapshot version we got from the first access of a table/version log), hidden version (all versions created after a client received its read version), write version (id of the created log/marker), and log version (table version during validation). Using those abstractions, we define conflict rules for our global version log. We consider a transaction as compatible if it does \textbf{not} fulfill any of the following rules:
\begin{itemize}
    \item[R1] read version $<$ hidden version - slack (outdated data)
    \item[R2] write version $\leq$ hidden version (log number already taken)
    \item[R3] log version + written logs $\not=$ write version - written logs (lost update)
\end{itemize}
In summary, all of our rules have different purposes: R1 ensures that a read query does not access data that is too outdated from the current state. This can be configured by each transaction via the parameter 'slack'. R2 ensures that no log version is assigned twice, handles concurrency, and commits during transaction execution. Lastly, R3 guarantees consistency by only committing logs that do not ignore prior changes. 
If a transaction fulfils no rule, it is compatible. In this case, it executes the table-level commit of each accessed table before atomically announcing the new snapshots. In LV[I]'s snapshot isolation mode, only R2 and R3 are enforced, making LV[I] only check write sets.

In our example in Figure \ref{fig:technique4}, T\textsubscript{4} commits first and updates the global version log. It advanced Bank x to version 8 and created Bank z. Afterwards, T\textsubscript{3} enters verification and recognizes that an unobserved change to Bank x happened (R3 fulfilled). Hence, T\textsubscript{3} must abort.

\labeltitle{LV[I] combines \#4.1 and \#4.2}
and implements it into the OTF structure. Module \#4.1 replaces parts of the Local Version Assignment and extends the Log Placement with Module \#4.2.
We define our LV[I] version log using the well-proven read-and-write set abstraction \cite{journals/tods/BernsteinG83}. However, in LHs, we deal with table and file versions that must not fulfill rules R1 -- R3. 
For transactions to retrieve globally consistent versions, they read the global version log and skip the snapshot assignment from their original OTF protocol. This change effectively reduces the number of requests to the metadata layers of single tables and replaces expensive requests (\eg LIST for Delta Lake \cite{journals/pvldb/ArmbrustDPXZ0YM20}). 

\labeltitle{LV[I]'s Read sets.}
LV[I]'s read sets contain all tables read by a transaction. Hence, the LV[CT] transaction does not update such table versions. During its execution, we implement checking for concurrent changes in each table and updating hidden versions. In the serializable protocol, reads must pass the validation phase. R1 in particular was designed to uphold a configurable distance between the read and hidden versions. However, LV[I] can be relaxed to gain write serializability by automatically letting each read-only transaction pass validation and only incorporating the hidden version if necessary. 

\labeltitle{LV[I]'s Write sets.} 
Similar to read operations, writes check for the hidden version of a table. A transaction's write set contains all tables where it effectively created at least one temporary log entry. The respective validation checks its changes and atomically updates the global version log.
To guarantee consistency, the transactions targeted write version number must be larger than their current log version, and the difference must match the number of added log files. LV[I] defines two special cases for writes in its validation: Updates and blind appends. Updates have to be read and written on the same table. Hence, they will have the same table in their read-and-write set. LV[I] enforces R1 -- R3 on those sets independently and concurrently, making them the most work-intensive operations.
On the other hand, blind appends only insert new data and are not bound by any restrictions. Hence, their order of operations is irrelevant. LV[I] implements a special flag for such operations that allows the transaction to take the following valid log ID directly in case of a validation failure.

\labeltitle{Constructing an LV[I] version log on demand.}
When accessing an OTF environment with LV[I] for the first time, the global version log does not exist. Hence, we construct this layer on demand using a mixture of the original OTF and module \#4: If an LV[I] client recognizes that no global version log exists, it proceeds by individually deducing its version of each accessed table. When committing, the transaction atomically enters the validation phase, accesses the empty global version log, and generates all entries based on its accessed table version. Other transactions accessing tables not contained in the version log will perform a similar mixture of deducing the version of the respective table themselves and adding the table during their validation. LV[I] defines adding a previously unknown table as always compatible because it can not check R1 -- R3. When executed for a while, this procedure generates the global version log without disrupting other clients and replaces the version assignment proposed by module \#4.

\labeltitle{LV[I]'s compatibility with other modules and original protocols.}
When using LV[I] alongside other client configurations, we face the problem of them ignoring the global version log. We ensure the compatibility of LV[I] by defining different behaviors for read-only and read-write/write-only transactions: Read-only transactions can still operate as described for module \#4. Such transactions do not conflict with concurrent transactions as they do not advance a table's version number. In contrast, all other transactions must be aware of such changes, as concurrent non-LV[I] read-writes/write-only transactions advance the version number without notifying the version log. If an LV[I] transaction decides to write to a table, it will check if the next possible version is available. If it detects a snapshot/log/marker of the same version, an inconsistency with the global version log occurs. It proceeds by reading the previously unknown logs, setting its hidden version to the newly read ID, and writing its results using the hidden version. During its commit, it counts the unknown log entries as it changes and commits them to the global version log if its initially assigned read version still equals the global version log. Otherwise, another concurrent transaction has already been committed using this version, and we must abort. Finally, to make read-writes/write-only transactions using LV[I] visible to all clients, LV[I] still generates a valid log entry for each table.

\section{The Guarantees of LakeVilla}
\label{sec:Guarantees}
After introducing the design of all LV features, we showcase two different aspects of LV: First, we showcase that LV[CT] will never block transactions indefinitely and resolves conflicts. Then, we explore the ACID guarantees achievable by prescribed LV combinations.

\subsection{LV[CT]'s Conflict Resolution}

\labeltitle{Proof Sketch: Hypothesis.} 
LV[CT] effectively exposes and resolves conflicts across tables without deadlocks by combining modules \#1 and \#3. For this proof, we assume that any marker dependency information is immediately available and DAG creation does not introduce overhead. Afterward, we explain how we resolve the practical overhead.

\begin{definition} 
\label{gen_def} We use the following definitions:

\begin{itemize}
    \item $Tbl$: The set of all unique tables
    \item $op_i$: A single OTF operation (e.g., read, write, ...) affecting a single table, $i \in \mathbb{N}$ 
    \item $T_x = (op_1, ..., op_n)$: A OTF transaction consisting of operations, $x, i \in \mathbb{N}$.
    \item $T = \{T_0, T_1, ...\}$: The set of all OTF operations
    \item $r_{\{table, T_x\}}$: The read snapshot by transaction $T_x \in T$ for $table \in Tbl$.
    \item $w_{\{table, T_x\}}$: The snapshot created during commit by transaction $T_x \in T$ for $table \in Tbl$.
    \item $tr : T \rightarrow (table_1, ..., table_n), table_i \in Tbl, i \leq n, n \in \mathbb{N}$ are all tables accessed by a transaction in $T$\\
\end{itemize}

\end{definition}

\labeltitle{Proof Sketch: Direct Proof.} 
Assume there are two concurrent transactions $T_1, T_2 \in T$ that both access at least one table during their execution:
\begin{equation}
    |tr(T_1) \cap tr(T_2)| > 0, T_1, T_2 \in T
\end{equation}
Following LV[CT], both transactions place a marker in the accessed tables and read the most recent snapshots.
From LV[R] (see Section \ref{sec:LakeVilla-Prototype}), we know that markers uniquely represent a running transaction within a table. Hence, their reserved write snapshot is always distinct: 
\begin{equation}
    \label{eq:unqiueness}
    \forall t \in tr(T_1) \cap tr(T_2): w_{(t, T_1)} \not= w_{(t, T_2)} 
\end{equation}
As LV[CT] transactions wait for preceding markers according to their local DAG, deadlocks can only occur if transactions form a circular wait scenario across tables, which first appears if an added edge would close a cycle in the DAG. W.l.o.g, assume $T_2$ accesses a table $t \in |tr(T_1) \cap tr(T_2)|$ after $T_1$, adding the edge $T_2 \rightarrow T_1$. Here, we have two scenarios:
\labeltitle{Case 1: $T_2 \rightarrow T_1$ closes a cycle.} Hence, there exists a path from $T_1$ to $T_2$ in the DAG. By adding its marker in $t$, $T_2$ creates the edge $T_2 \rightarrow T_1$, forming a cycle in the DAG. $T_2$ checks $T_1$'s marker in $t$ and detects that $T_2$ is listed as $T_1$'s dependency. Hence, it detects the conflict and triggers a marker shift.  

Performing a marker shift (module \#3) effectively frees all markers of $T_2$ within such a cycle and claims newer ones. Hence, all previous outgoing DAG edges between the transaction performing the shift and all waiting transactions are inverted for the respective table. Therefore, during their concurrent checks, waiting transactions will realize that the marker is freed and proceed, breaking the cycle. We further know that the result must be acyclic. If we add the inverted edge $T_2 \leftarrow T_1$ and it closes a cycle, we know that a path from $T_2$ to $T_1$ already exists. Combined with the previously found path from $T_1$ to $T_2$, we know that the DAG contained a cycle, which violates DAG constraints. Hence, a marker shift breaks any new cycle by adding an inverted edge  $T_2 \leftarrow T_1$. We summarize this observation as:

\begin{lemma}
\label{marker_shift}
    Given a cycle $C$ in the dependency DAG, a marker shift on $T_i \in C = (T_a, T_b, ...)$ inverts all incoming edges of $T_x$, guaranteeing:\\
    $\forall c \in C/{T_x}, \forall s \in tr(c) \cap tr(T_x): w(s, c) < w(s, T_x)$
\end{lemma}

\labeltitle{Case 2: $T_2 \rightarrow T_1$ does not close a cycle.}
Trivial, the resulting DAG is still acyclic, and $T_2$ can continue.

\labeltitle{Proof Sketch: Conclusion.} Hence, LV[CT] does not resolve in deadlocks during execution, given that the dependency DAG is complete. \qed

\labeltitle{Handling Delays and Overhead in Practice.} Due to the latency of object stores and to save costs, a client might not have the entire information available to build the entire DAG. Hence, it might be possible that larger cycles exist that remain undetected. For such cases, we implemented a timeout strategy for markers: A marker is guaranteed to only exist for a specified time. Further, if a transaction waits too long for another transaction, it preemptively triggers a marker shift, breaking potential larger cycles. In our LH use case, this mixture proved effective where LV[CT]'s module \#3 resolves smaller cycles between transactions. In the rare case of concurrent, long-running transactions, marker timeouts prevent deadlocks and limit the DAG building overhead.

\subsection{ACID Guarantees}

\begin{table*}[t]
\small
\caption{Highest ACID guarantees achievable for OTFs by enforcing specific LV features}
\label{fig:acid-overview}
\centering
\begin{tabular}{l|llll}
 & \makecell[ll]{Iceberg/Delta Lake} & \makecell[ll]{LakeVilla{[}Recovery{]}} & \makecell[ll]{LakeVilla{[}Complex Txn.{]}} & \makecell[ll]{LakeVilla{[}Isolation{]}} \\ \hline \makecell[cc]{\textbf{A}} & \makecell[ll]{OTF structure \textbf{(ST)}} & \makecell[ll]{Atomic Markers \textbf{(ST)}} & \makecell[ll]{Atomic Markers \textbf{(ST)}} & \makecell[ll]{Version Validation}\\\hline
    \makecell[cc]{\textbf{C}}  & \makecell[ll]{Eventual \textbf{(SQ,SW)}} & \makecell[ll]{Causal \textbf{(ST)}} & \makecell[ll]{Causal} & \makecell[ll]{Linearizability} \\\hline
    \makecell[cc]{\textbf{I}}   & \makecell[ll]{Serializable \textbf{(SQ,ST)} \cite{isolation_databricks,conf/cidr/0001KPDSZ23}} & \makecell[ll]{Serializable \textbf{(ST)}} & \makecell[ll]{Repeatable reads} & \makecell[ll]{Serializable}\\\hline
    \makecell[cc]{\textbf{D}}  & \makecell[ll]{Object store guarantee} & \makecell[ll]{Object store guarantee} & \makecell[ll]{Object store guarantee} & \makecell[ll]{Object store guarantee} \\ \hline
\end{tabular}

\medskip
\vspace{0.1cm}
\small
Single Query only: \textbf{(SQ)} \quad Single Table only: \textbf{(ST)} \quad Single Writer only: \textbf{(SW)}
\end{table*}

LV allows transactions to choose specific features by temporarily implementing the requested LV feature. However, architects can build LHs with different ACID properties by enforcing different features on all transactions. We now explore these possibilities and prove their effectiveness. Our findings are summarized in Table \ref{fig:acid-overview}.

\subsubsection{\textbf{A}tomicity of LakeVilla}
\label{atomicity}
All LV Features guarantee table-level atomic transactions as shown in Table \ref{fig:acid-overview}. Like the base OTFs, their Atomicity relies on atomic object store operations (if available), concurrent checks (see \cite{journals/pvldb/ArmbrustDPXZ0YM20}), or strong consistency with PUT-IF-ABSENT operations \cite{aws_conditional_writes}.

For LV[R] and LV[CT], Atomicity revolves around their usage of atomic markers. Given the metadata structure of an OTF table, a transaction overwrites its marker with all changes made during its commit. Hence, any concurrent transaction will see either the marker (see Section \ref{subsec:LV-R}) or the complete log file. Because a marker is a single file and each marker reserves a unique write snapshot for its transaction (see Equation \ref{eq:unqiueness}), all changes of a transaction become visible at the same time when overwritten. However, this is limited to single tables due to the district metadata layers of OTF tables.

In contrast, LV[I] guarantees Atomicity through its validation step, which uses atomic operations to allow only one client to modify the global version log. The head of the global version log is a single file that can be overwritten by a validated transaction in one operation, globally exposing all its changes simultaneously.

\subsubsection{\textbf{C}onsistency of LakeVilla}

This section discusses the database and distributed consistency levels of LV. It uses Definition \ref{gen_def} and extends it by:
\begin{definition}
\label{order_def}
    We describe the ordering between two operations $op_1, op_2$ using:
    \begin{itemize}
        \item $op_1<op_2$: $op_1$ before $op_2$
        \item $op_1>op_2$: $op_1$ after $op_2$
        \item $op_1=op_2$: $op_1$ is concurrent to $op_2$
        \item $op_1\sim op_2$: $\sim \in \{<, >, =\}$
        \item $op_1\not\sim op_2$ no order between $op_1$ and $op_2$
    \end{itemize}
    Equivalently, for two transactions $T_1, T_2 \in T$.
\end{definition}

\labeltitle{The Database Consistency of LV} depends on the base OTF. Due to their metadata structure, Delta Lake and Iceberg offer ACID transactions for single tables \cite{journals/pvldb/ArmbrustDPXZ0YM20,Iceberg_website,delta_doc}. All OTF transactions are ordered by the table's "chain" or "tree" of the respective OTF metadata layer. Due to the atomic operations used (PUT-IF-ABSENT, Conditional Writes, extra service), each transaction claims a distinct spot in the metadata layer and places all operations of a transaction in a single log file - defining the order of all successful transactions and their operations within a table $x$.
\begin{lemma}
    \label{base_guarantee}
    OTFs guarantee an ordering of all transactions and their operations on the same table:\\
    $\forall T_1, T_2 \in T: tr(T_1) \cap tr(T_2) \not= \emptyset \Leftrightarrow T_1 \sim T_2 \Leftrightarrow \forall op_1 \in T_1, op_2 \in T_2: op_1 \sim op_2$
\end{lemma}
Because LV's features keep the OTF design and always lead to a valid state of the base OTF, all its features guarantee table-level consistency. 

\labeltitle{LV[R]: Causal Consistency (Single-Table).}
The causal consistency of LV[R] among clients using this feature is implied by its atomic markers and transactions, which always publish all their operations in a single file. Combined with the existing ordering mechanisms of OTFs (Lemma \ref{base_guarantee}), they guarantee a logical and consistent ordering between transactions and their operations.

\labeltitle{Proof Sketch: Hypothesis.} LV[R] guarantees causal consistency for single-table, multi-query transactions.

\labeltitle{Proof Sketch: Direct Proof} Following Definition \ref{gen_def}, all transactions $T_i\in T$ define an order among their operations. Hence:
\begin{equation}
\label{ops_order}
    \forall T_i \in T, \forall x,y \in T_i, x \not = y: x \sim y
\end{equation}
LV[R]'s marker technique guarantees that all $ops \in T_i \in T$ affecting the same table are placed in the same metadata file (Section \ref{atomicity}). Following Lemma \ref{base_guarantee}, the OTF layer defines an order between transactions on the same table (e.g., in Delta Lake: log files). Thus:
 \begin{equation}
 \label{txn_order}
     \forall T_1, T_2 \in T, 
     |tr(T_1)| = 1,  tr(T_1) = tr(T_2): T_1 \sim T_2
 \end{equation}
As LV[R] is applied to the base OTF, combining Equation \ref{ops_order} and Equation \ref{txn_order} gives:
\begin{lemma}
\label{lvr-consistency}
LV[R] guarantees causal consistency on a table level using its marker technique:\\
    $\forall T_1, T_2 \in T, |tr(T_1)| = 1,  tr(T_1) = tr(T_2), x \in T_1, y \in T_2, x\not=y: x \sim y$
\end{lemma}

\labeltitle{Proof Sketch: Conclusion.} This implies table-level causal consistency for LV[R] as only the order of related operations is defined. This result was expected as LV[R] focuses on table-level guarantees and does not track dependencies between tables like the base OTFs \qed

\labeltitle{LV[CT]: Causal Consistency.}
LV[CT] provides causal consistency globally. A client's operations are ordered by their execution order (Equation \ref{ops_order}). Concurrent operations accessing the same table(s) are also related because they access the same data items ($|tr(T_1) \cap tr(T_2)| \not= 0,$). We define the order of such operations within a table using atomic markers like LV[R] (compare Equation \ref{txn_order}). 
Unlike LV[R], LV[CT]'s conflict detection builds a DAG of dependencies for the respective transaction, guaranteeing a consistent global ordering between such transactions and their operations, lifting the restriction to single-table transactions: 
\begin{lemma}
\label{lvct-consistency}
LV[CT] guarantees causal consistency globally using its marker and conflict detection techniques: \\
    $\forall T_1, T_2 \in T, |tr(T_1) \cap tr(T_2)| \not= 0, x \in T_1, y \in T_2, x\not=y: x \sim y$
\end{lemma}
However, LV[CT] is not serializable because it does not define an order among operations on distinct table sets. 

\labeltitle{LV[I]: Linearizability.}
Regarding LV[I]'s isolation, we designed the validation process to be atomic. Hence, all write transactions must pass the validation sequentially, defining a global order similarly to LV[R]'s marker on table level. However, overlapping read-only transactions that retrieve their versions from the global version log during an active validation phase will work with outdated data. Logically, when read-only transactions enter validation, they will likely succeed depending on the interval chosen for LV[I]'s R1. Hence, the read-only transaction must be ordered before the transaction that previously processed the validation. This guarantees a serial ordering between all transactions and, thus, Linearizability.

\begin{lemma}
\label{lvi-consistency}
LV[I] guarantees Linearizability globally using its global version log techniques: \\
    $\forall T_1, T_2 \in T, x \in T_1, y \in T_2, x\not=y: x \sim y$
\end{lemma}

\subsubsection{\textbf{I}solation}

\labeltitle{Base OTFs: Serializability (Single-Query, Single-Table).} 
OTF formats guarantee serializability limited to single-query and single-table transactions \cite{journals/pvldb/ArmbrustDPXZ0YM20}. Since OTFs only feature single-query and single-table transactions, all transaction operations are contained within a single log file, providing serializability or snapshot isolation \cite{isolation_databricks,conf/cidr/0001KPDSZ23}. As a result, the execution order of all transactions equals a serial execution. Their operations can be ordered transitively to form a serial history between all operations on the same table. 

\begin{lemma}
    \label{base_guarantee2}
    OTF guarantee a serial ordering of transactions and their operations:\\
    $T_1, T_2 \in T, tr(T_1) = tr(T_2), |T_1| = |T_2| = 1: T_1 < T_2 \Rightarrow \forall op_1 \in T_1, op_2 \in T_2: op_1 < op_2$
\end{lemma}

\labeltitle{LV[R]: Serializability for single tables.}
LV[R] extends Lemma \ref{base_guarantee2} by adding atomic markers placed during a transaction's start to put all transaction operations into a single file. Hence, all distinct transactions operating on the same table initially get a unique version and are ordered by the base OTF structure (Lemma \ref{base_guarantee2}).

\labeltitle{Proof Sketch: Hypothesis.} LV[R] guarantees serializability for single-table transactions.

\labeltitle{Proof Sketch: Proof by contradiction.} Assume LV[R] produces a non-serializable history for all single-table, multi-query transactions $T$. Then, at least two transactions $T_1, T_2 \in T$ can not be ordered. 
Since both transactions operate exclusively on the same table (single-table transactions), they must add their operations to the same metadata later. Because $T_1 \not\sim T_2$ and OTFs always define an order of operations, we know from Lemma \ref{base_guarantee2} that there exists a circular dependency:
\begin{equation}
    \label{cyclic3}
    \exists a, b \in T_1, \exists x, y \in T_2: a < x \cap b > y
\end{equation}
However, the structure of OTFs (compare section \ref{sec:metadata_design}) implies that at least one transaction must have written multiple files containing distinct operations.

\labeltitle{Proof Sketch: Conclusion.} This is a direct violation of LV[R] commit protocol (overwrite a marker with a log file containing all operations). Hence, different log files must belong to different transactions, and Equation \ref{cyclic3} does not apply. Hence:

\begin{lemma}
    \label{lvr-isolation}
    LV[R] guarantees serializability for multi-query, single-table transactions using its marker technique:\\
    $T_1, T_2 \in T, tr(T_1) = tr(T_2), |tr(T_1)| = 1: T_1 < T_2 \Rightarrow \forall op_1 \in T_1, op_2 \in T_2: op_1 < op_2$
\end{lemma}

\begin{table*}[bt]
\caption{Shortened Hermitage Results (multi-query) for Spark using Delta Lake (Spark[DL]) or Iceberg (Spark[IB]) and LakeVilla (LV). Anomalies described with "-MT" are multi-table versions of the original scenario. We tested all anomalies and show the ones with the biggest differences here}
\small
\label{fig:hermitage}
\centering
\begin{tabular}{l|lllll}
   Anomaly & \makecell[ll]{Spark{[}DL{]}} & \makecell[ll]{Spark{[}IB{]}}& \makecell[ll]{LV{[}R{]}}& \makecell[ll]{LV{[}R,CT{]}}& \makecell[ll]{LV{[}R,CT,I{]}}\\ \hline
   \makecell[ll]{\textbf{G0-MT} (Write Cycles)} & \greenCheck* & \greenCheck* & \No & \greenCheck & \greenCheck\\ \hline
    \makecell[ll]{\textbf{G1a} (Aborted Reads)} & \No & \No & \greenCheck & \greenCheck & \greenCheck \\ \hline
     \makecell[ll]{\textbf{G1b} (Interm. Reads)} & \No & \No  & \greenCheck & \greenCheck & \greenCheck\\ \hline
     \makecell[ll]{\textbf{G1c-MT} (Circ. information flow)} & \No* & \No* & \No & \greenCheck & \greenCheck\\\hline
     \makecell[ll]{\textbf{OTV} (Obs. Txn. Vanishes)} & \No & \No & \greenCheck& \greenCheck& \greenCheck\\\hline
     \makecell[ll]{\textbf{P4} (Lost Update)} & \No & \No & \greenCheck& \greenCheck& \greenCheck\\\hline
     \makecell[ll]{\textbf{Concurrent Reads} (Linearizability)} & \textbf{-} & \textbf{-}& \No& \No& \greenCheck \\
\end{tabular}

\medskip
\vspace{0.1cm}
\small
prevents anomaly: \greenCheck; not prevented: \No.\\ *=implied by multi-query results and limitations
\end{table*}

\labeltitle{LV[CT]: Repeatable Reads.}
LV[CT] removes Lemman\ref{base_guarantee2}'s restriction on single tables ($|tr(T_1)| = 1$). Because it reuses the marker module (\#1), we only prove that LV[CT] ensures that it creates a serializable order of transactions across tables. All other proofs are analogous to LV[R]'s for operations on the same table.

\labeltitle{Proof Sketch: Hypothesis.} LV[CT] provides repeatable reads across tables for multi-table, multi-query transactions.

\labeltitle{Proof Sketch: Direct Proof.} LV[CT] sets markers in each accessed table of a transaction. Hence, it reserves a set of table versions within the respective metadata layers. Hence: 

\begin{equation}
    \forall T_i \in T, \forall tbl \in tr(T_i), \exists w_{(tbl, T_i)} \in \mathbb{N}
\end{equation}

\noindent
By applying Lemma \ref{lvr-isolation} per table in $T$, we see that this orders all operations for each table.
However, unlike LV[R], if a transaction accesses a new table, LV[CT] sets a new marker in the respective metadata layer. Given two concurrent multi-table transactions $T_1, T_2 \in T$, $T_2$ decides to access a new table $tbl$. In this scenario, we have two possibilities:

\labeltitle{Case 1: $T_1$ did not access $tbl$ yet.} In this case, $T_1$ executed no operations on $tbl$ beforehand. Hence, all of $T_1$'s operations can be ordered before or after $T_2$'s marker version on $tbl$, depending on the existing relation $T_1 \sim T_2$ on another table. Hence, $T_1$ is isolated by default. There is no need for LV[CT] to readjust this situation.

\labeltitle{Case 2: $T_1$ already accessed $tbl$.} Then, a marker for $T_1$ in $tbl$ already exists. When $T_2$ sets a new marker, this automatically forms the relation:
\begin{equation}
    T_1 < T_2 \Leftrightarrow w_{(tbl, T_1)} < w_{(tbl, T_2)}
\end{equation}
If there are previously executed operations in $T_1$ and $T_2$, we have three scenarios that may apply here:
\labeltitle{Case 2.1: $(tr(T_1) \cap tr(T_2) = \{tbl\}$.} Hence, there is no other table that $T_1$ and $T_2$ have both accessed yet. Their global relation is updated to $T_1 < T_2$, and both can continue. On commit, all of $T_1$'s operations are logically ordered before $T_2$. Hence, $T_2$ must compare its changes to the respective logs and either commit or abort.

\labeltitle{Case 2.2: $\forall z \in (tr(T_1) \cap tr(T_2)/tbl, w_{(z, T_1)} < w_{(z, T_2)}$.} In this scenario, both transactions encountered each other in another table beforehand (If they are empty, case 2.1 applies).
In these tables, $T_1$ and $T_2$ already have the newly created relation within $tbl$ globally. Thus, LV[CT] automatically upholds this relation, and both transactions can continue normally.  
 
\labeltitle{Case 2.3: $\forall z \in (tr(T_1) \cap tr(T_2)/tbl, w_{(z, T_1)} > w_{(z, T_2)}$}
Hence, LV[CT] observes two contradicting relations in different tables and triggers a marker shift. As LV[CT] can not rearrange the order within $tbl$, the marker shift frees all markers in all other tables and places a new marker. This effectively inverts the relation in such tables as described in Lemma \ref{marker_shift}.
Hence, with the relation found in $tbl$ and after a maker shift, LV[CT] guarantees: 
\begin{equation}
    \forall z \in tr(T_1) \cap tr(T_2), w_{(z, T_1)} < w_{(z, T_2)}
\end{equation}
This fixes the global relation between those two transactions. 
Since Cases 2.1, 2.2, and 2.3 ensured consistent ordering between transactions, LV[CT] provides an order among the tables. 

However, LV[CT] is not serializable due to the marker shift: A transaction can never read dirty data due to the marker replacement during the commit. To prevent Non-repeatable reads, LV[CT] 's marker shift never changes the read version of each table $r$. Despite a compatibility check during commit, phantom reads can occur: 
If a transaction requests a range of tuples, it usually accesses multiple data files in the OTF table. As LV[CT] only remembers previous reads, it can not predict any changes incorporated by a marker shift (if a transaction that was previously dependent and now became preceding) belongs to a range query. Hence, for LV[CT], the same query might look like a different operation that will work with the new changes incorporated.

\labeltitle{Proof Sketch: Conclusion.} 
Because LV[CT] defines a structure between dependent transactions in all scenarios but might incur phantom reads, it provides repeatable reads globally.\qed

\begin{lemma}
    \label{lvct-isolation}
    LV[CT] guarantees repeatable reads globally using its conflict detection technique
\end{lemma}

\labeltitle{LV[I]: Serializability.}
LV[I] provides the isolation level of serializability. Its global version log defines a total order between all committed transactions. Each transaction must enter its atomic validation phase and check its changes using R1 -- R3.

\labeltitle{Proof Sketch: Hypothesis.} LV[I] guarantees serializability 
\labeltitle{Proof Sketch: Proof by contradiction.} Assume LV[I] is not serializable. Then, two transactions $T_1, T_2 \in T$ can not be ordered sequentially ($T_1 \not\sim T_2$). 
Since LV[I] takes version assignment to its global version log, both transactions received their read and write versions. Since $T_1$ is able to read $T_2$'s data, we know:
\begin{equation}
    \label{p1}
    \exists tbl \in tr(t_1) \cap tr(t_2): w_{(tbl, T_2)} < r_{(tbl, T_1)}
\end{equation}
Thus, the global version log must have previously known $w_{(tbl, T_2)}$. Hence, $T_2$ must have passed the validation phase, followed by committing in all its accessed tables before making its changes globally available. However, $T_2$ observed $T_1$'s changes, implying:
\begin{equation}
    \label{p2}
    \exists tbl_2 \in tr(T_1) \cap tr(T_2): r_{(tbl_2, T_2)} > w_{(tbl_2, T_1)}
\end{equation}
Hence, $T_2$ received a version $r_{(tbl_2, T_2)}$ larger than $w_{(tbl_2, T_1)}$ from the global version log. Hence, $T_1$ must have passed validation and committed before $T_2$. However, equations \ref{p1} and \ref{p2} are impossible to fulfill at the same time with the atomic validation phase of LV[I]. Transactions pass through it sequentially and end it by committing or aborting. Hence, there can not be a circular dependency between $T_1$ and $T_2$.

\labeltitle{Proof Sketch: Conclusion.}
Due to the atomic design of the global validation phase, LV[I] guarantees that previously validated operations will be ordered before and after the current transaction. Similarly, as the global version log is a global layer for the entire LH, it defines relations between all transactions, giving Serializability.\qed

\begin{lemma}
    \label{lvi-isolation}
    LV[I] guarantees serializability globally using its global version log:\\
    $T_1, T_2 \in T: T_1 < T_2 \Rightarrow \forall op_1 \in T_1, op_2 \in T_2: op_1 < op_2$
\end{lemma}

Moreover, if we relax R1 using the 'slack' parameter to a larger area to let read-only transactions pass validation without a check, we skip the check on the transactions' read sets.

\begin{lemma}
\label{lv-i-isolation2}
LV[I] with a relaxed R1 guarantees snapshot isolation globally.
\end{lemma}

\subsubsection{\textbf{D}urability of LakeVilla.}

Similar to the original OTF protocols, the guarantees of object stores and a client successfully uploading all new files before generating the log file (\eg compare \cite{journals/pvldb/ArmbrustDPXZ0YM20}) achieve Durability for LV[R] and LV[CT]. Further, a transaction only updates LV[I]'s global version log if it successfully enters validation and all tables' data files are written. Hence, LV[I] uploads all data files to the object store before committing and can rely on the object store guarantees to provide Durability.

\subsubsection{Experimental Validation}

To provide further validation of LV's guarantees and show the combined effect of its features, we experimentally check Spark and our prototype using LV[R], LV[R, CT], and LV[R, CT, I] for the anomalies described by Hermitage \cite{hermitage}. Table \ref{fig:hermitage} shows a selection of the results with additional multi-table versions of the same anomaly (marked with -MT). The original Delta Lake (Spark[DL]) and Iceberg (Spark[IB]) were included despite their restriction on single-query transactions, which led them to fail most multi-query anomalies. 
An exception is G0-MT (write cycles), avoided by their limitations and thus marked uniquely (G0-MT). In contrast, LV[R] prevents table-level anomalies like lost updates (P4), aborted reads (G1a), or Intermediate Reads (G1b) and adapts to aborted/crashed transactions (OTV). When adding LV[CT], the design can resolve conflicts across tables like Write Cycles (G0-MT) and Circular information flow (G1c-MT). Further, LV[I]'s global isolation guarantees concurrent transactions to observe globally consistent versions across tables. In summary, these results confirm our targeted guarantees. 

\section{Evaluation}
\label{sec:evaluation}

\labeltitle{Overview.} We start by analyzing the different phases of the LV Features and comparing them to Spark's Delta Lake \cite{spark_web,delta_lake_web}. We investigate their effects under concurrent workloads using YCSB\footnote{YCSB-LH available: \url{https://github.com/goetztj/YCSB-LH}} \cite{conf/cloud/CooperSTRS10,YCSB-C} and test different feature combinations based on the cloud analytics benchmark (CAB) \cite{journals/pvldb/RenenL23}. We conclude by running LV on selected analytical queries from TPC-DS, testing for long-term effects with LSTBench, and estimating the added overhead for systems like Spark.

\subsection{Experimental Setup}
\labeltitle{Object Store: AWS S3 and MinIO.} For all experiments, we either use AWS S3~\cite{s3_web} or MinIO~\cite{minioWeb} on a cloud instance with an Intel(R) Xeon(R) Platinum 8260 CPU with 48 cores at 2.40GHz with 524GB of RAM and 5TB of SSD storage running Linux. 
S3 and MinIO provide strict consistency \cite{minio_str_consistency,aws_str_consistency} and conditional writes \cite{aws_conditional_writes}. Further, MinIO offers analytical features for the experiments. Both services are accessed through the AWS S3 SDK. An experiment uses S3 is not stated differently.

\labeltitle{LakeVilla Prototype.}
Our C++ LV prototype\footnote{LakeVilla prototype available: \url{https://github.com/goetztj/LakeVilla}.} implements a baseline Delta Lake client (LV[DL]) with all modular features (LV[R], LV[CT], LV[I]).
It runs on an Intel(R) Xeon(R) Platinum 8260 CPU at 2.40GHz with two sockets of 64 cores and 1048GB RAM.
We chose eight threads per LV client unless stated otherwise.

\labeltitle{Spark: Delta Lake.} Spark 3.5.3 runs with Hadoop 2.12, Delta Lake 3.0.0 (checkpoint interval 10), and Hadoop AWS 3.3.4.
We minimized caching to observe the OTFs' effects on the system. 
Our HIVE 3.0.0 catalog and its PostgreSQL backend run in two separate docker containers. 
Spark uses Java 11 and runs distributed on 9 machines with the same setup as above.
The coordinator runs on one server with specifications like our MinIO server and shares its resources with the catalog store and benchmark drivers to reduce the networking effects between these components.
For workers, we configured two instances per server using an Intel(R) Xeon(R) Platinum 8260 CPU at 2.40GHz with 8 cores each. Each worker used 64GB RAM and eight cores. 

\subsection{Phases of OTFs}

\begin{figure}
    \begin{minipage}{0.38\textwidth}
        \centering
        \begin{subfigure}[t]{\textwidth}
            \centering
            \includegraphics[width=0.9\textwidth]{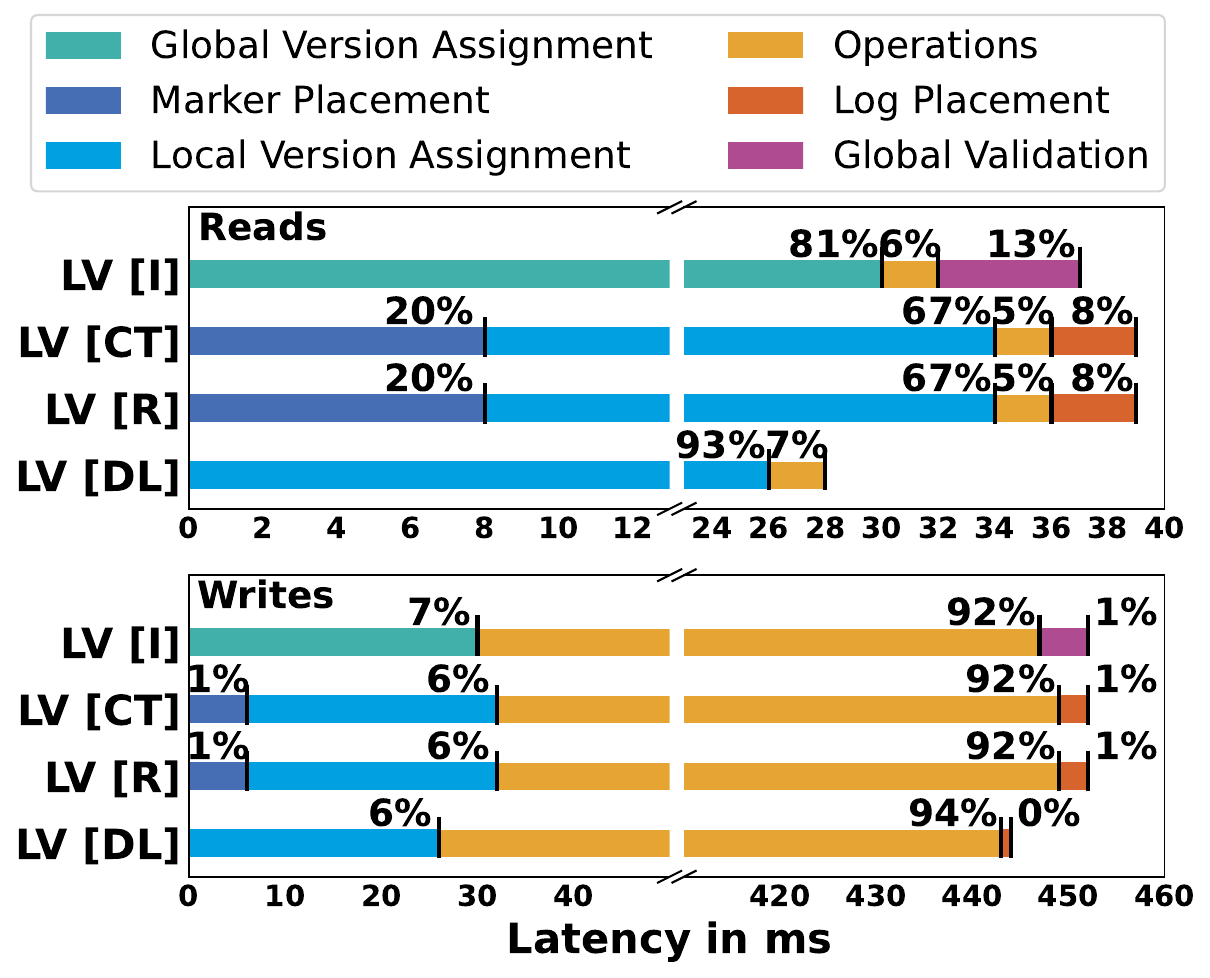}
            \caption{LakeVilla Prototype}
        \label{fig:Phases-LV}
        \end{subfigure}
    \end{minipage}
    \hfill
    \begin{minipage}{0.09\textwidth}
        \centering
        \begin{subfigure}{\textwidth}
            \centering
            \includegraphics[width=\textwidth]{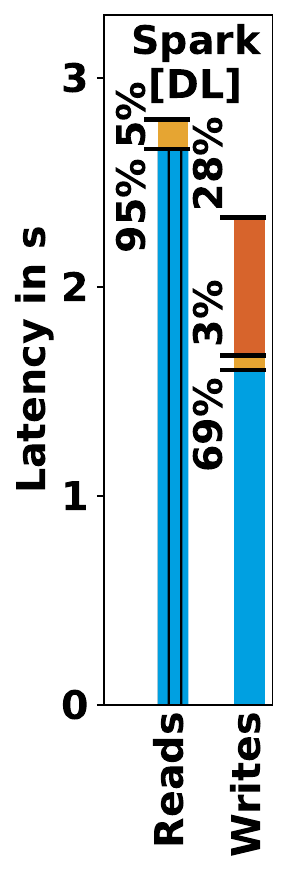}
            \caption{Spark}
            \label{fig:Phases-spark}
        \end{subfigure}
    \end{minipage}
     \caption{Phases for read and write workloads of a single transaction for YCSB queries on MinIO (matplotlib)}
        \label{fig:Phases}
\end{figure}

\labeltitle{Comparison of Spark and LakeVilla.}
We implemented our LV prototype as a separate C++ client. Thus, we first compare the differences between it and other systems. For this purpose, we ran YCSB sequentially with Spark using Delta Lake (Spark[DL]) and LV[DL]. If not stated otherwise, we use the default YCSB of 1000 operations on 1000 entries with row sizes of 1KB \cite{conf/cloud/CooperSTRS10}. We logged the different phases of both implementations and calculated both the average latency for reads and writes, and how much each phase contributes to the overall latency. We expect both systems to display similar behavior. We show the results in Figure \ref{fig:Phases}. Spark[DL] and LV[DL] indeed behave similarly for reads. 
They spend most of their time accessing and reading the metadata of the YCSB table (95\% and 93\%, respectively) and the rest to load the requested data.

We observe a notable difference for the write requests. Spark[DL] 's write latency is dominated by accessing the metadata and generating a new log entry within the metadata layer (69\% and 28\%). The actual operation of uploading new data is just a fraction of its latency. In contrast, LV[DL] 's numbers indicate more time spent on data generation, while access to the metadata layer and log generation play a minor role. This difference is critical to interpreting the subsequent results in our evaluation so we can reliably estimate the impact of integrating LV in systems like Spark.

\labeltitle{The different LV Features.}
Next, we evaluate the impact of each LV feature on latency. We use the same setup as before and show our results in Figure \ref{fig:Phases-LV}. First, we analyze the overhead of placing LV[R]'s and LV[CT]'s markers (module \#1). We hypothesize that it affects the latency, especially for reads. In Figure \ref{fig:Phases-LV}, writes show minimal impact as they align with metadata layer operations in the original OTF (LV[DL]). Further, freeing those markers is fused with the write transaction's log placement, compensating for its effect as seen in Figure \ref{fig:Phases-LV}. 
In contrast, reads experience a 20\% latency increase due to added OTF synchronization, with an additional impact when we free the markers post-read. This indicates that LV[R] and LV[CT] are better suited for write operations.

Next, we analyze the impact of LV[R] 's sublog (module \#2) and LV[CT]'s deadlock detection (module \#3) on latency using the same setup. Because both modules are expected to work concurrently alongside the standard operations, we expect minimal impact on latency, which is also confirmed by Figure \ref{fig:Phases-LV}. The time spent by the prototype on the actual operation in LV[R] and LV[CT] matches the baseline (LV[DL]). LV[R]'s sublog generation is distinct and does not impact the baseline operation. Further, LV[CT]'s deadlock detection does not detect problems as YCSB uses only one table. Hence, under non-concurrent workloads, neither module impacts read or write latency.

Lastly, we expect LV[I]'s global version log (module \#4) to have a noticeable but smaller impact on latency in the single-table YCSB setup. As seen in Figure \ref{fig:Phases-LV}, it replaces the base table's local version assignment, adding a 4ms overhead for reads and writes compared to the baseline. Further, even though the validation phase takes longer for writes, it more significantly affects the latency of the reads. In LV[I]'s WriteSerializable mode, version assignment is skipped for reads, thereby shortening their latency. In summary, the global version log introduces overhead for all operations in a sequential, single-table setting. However, the WriteSerializable mode of LV[I] can mitigate this impact in such scenarios.

\begin{figure}
    \centering
        \centering
        \includegraphics[width=0.45\textwidth]{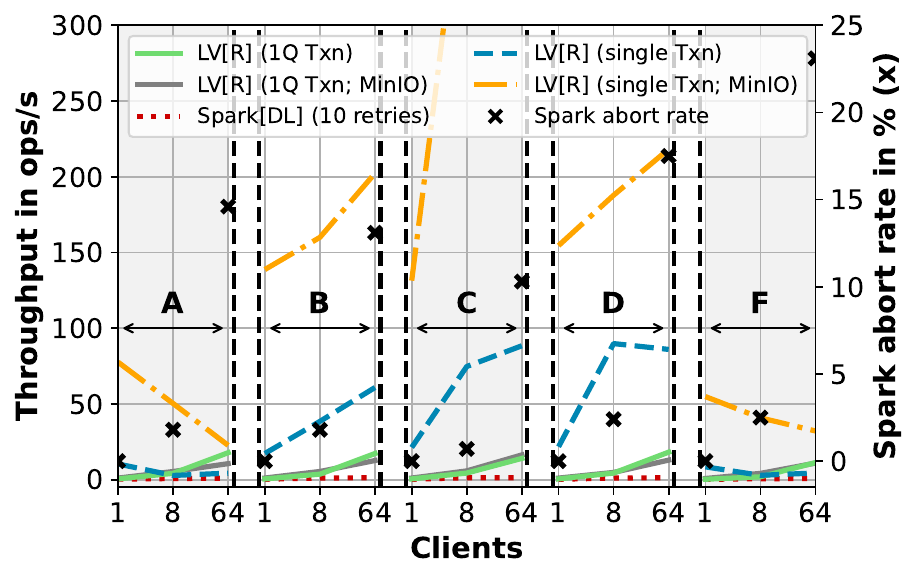}
    \caption{YCSB when using Spark (Delta Lake) and LV[R] on S3 and MinIO (matplotlib)}
    \label{fig:ycsbClientScale}
\end{figure}

\subsection{Concurrent Workloads}

\labeltitle{OTFs and Concurrent Workloads.} To evaluate the effects of concurrency and supporting multiple writers, we execute YCSB \cite{conf/cloud/CooperSTRS10,YCSB-C} with 1, 8, and 64 concurrent clients and compare Spark[DL] with LV[R] on S3 and MinIO. Spark was allowed up to 10 retries for a failed query. We evaluated LV[R] using single-query transactions (1Q Txn) and multi-query transactions that bundle all operations of a client (single Txn). We expect the throughput to increase for all workloads and systems, but also that the frequency of the conflicts will impact the behavior both through aborts by affecting the throughput. 
Figure \ref{fig:ycsbClientScale} shows our results for YCSB A, B, C, D, and F. YCSB E was not tested because our current LV prototype does not support scans. Additionally, we only show the results for Spark[DL] on S3 as they represent both scenarios equally.
Confirming our hypothesis, Spark experienced a slight increase in throughput, but with abort rates rising alongside the client count. Surprisingly, we also measured a small number of aborts for the read-only workload YCSB C, which we found to be an internal problem in Spark.
In contrast, LV[R] had no aborts and outperformed Spark in throughput, particularly when bundling operations, which reduces version assignments per client.
When we include the commit latency of LV[R] (single Txn) in our throughput calculations, we can also measure the impact for each workload type: For example, write-heavy workloads (A, F) experience a drop in throughput when we increase concurrency due to redo operations triggered at commit time. In contrast, executing LV[R] (1Q Txn) degenerates into normal Delta Lake with built-in retries, resulting in a throughput closer to Spark. As to be expected, bundling multiple operations into a transaction is advantageous for throughput as it minimizes the overhead of the version assignment phase. However, when overused, it may lead to penalties due to redo operations. We observed these results for both S3 and MinIO. In the S3 case, higher network latency decreased the overall throughput, mostly penalizing bundling operations into a single transaction. In conclusion, independent of the used object store, we show that concurrency for writes is possible with LV[I], but users must be mindful of the suitable transaction sizes. The more writes a workload has, the fewer requests should be bundled into single transactions.

\begin{figure}
    \centering
    \begin{subfigure}{.2\textwidth}
        \centering
        \includegraphics[height=4.2cm]{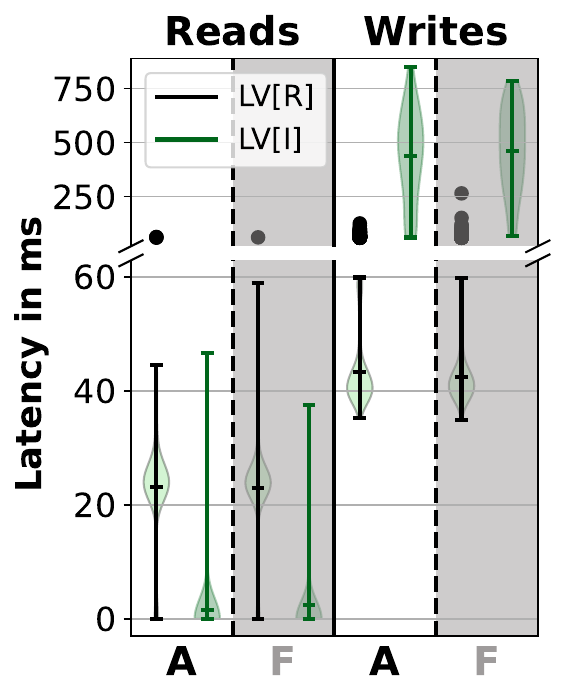}
        \caption{50\% reads}
        \label{fig:YCSB-comp-write-heavy}
    \end{subfigure}\hspace{-1.5em}
    \begin{subfigure}{.2\textwidth}
        \centering
        \includegraphics[height=4.2cm]{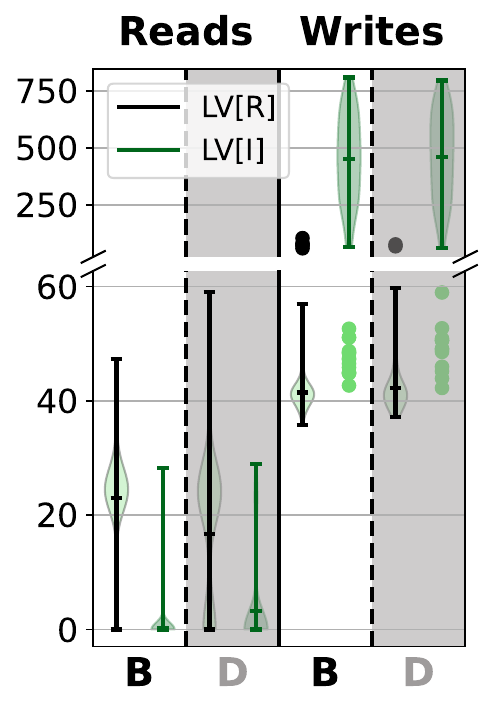}
        \caption{95\% reads}
    \label{fig:YCSB-comp-read-heavy}
    \end{subfigure}\hspace{-1.5em}
    \begin{subfigure}{.12\textwidth}
        \centering    \includegraphics[height=4.2cm]{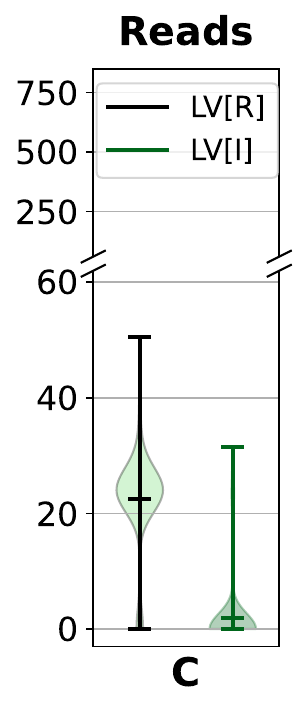}
        \caption{100\% reads}
    \label{fig:YCSB-comp-read-only}
    \end{subfigure}
    \caption{YCSB (8 clients, S3): Comparison of LV[I] and LV[R] (matplotlib)}
    \label{fig:YCSB-comp}
\end{figure}

\labeltitle{LV[R] and LV[I].}
Lastly, we use YCSB with varying workload types to compare the impact of LV[R] 's markers and LV[I] 's global version log on latency under concurrency. Again, we bundled all operations into one transaction per client and ran YCSB with eight clients using S3. LV[I] 's hidden version interval was set to two.
We hypothesize that LV[R] will be more suitable for mixed workloads, whereas LV[I] will be better for read-heavy workloads. In Figure \ref{fig:YCSB-comp}, we report this experiment's read and write latencies. They confirm our hypothesis in the following way: LV[R] provides better write latency at the cost of more outliers. These are caused by marker acquisition failures, clean-up, and transaction retries. LV[R] adds the same overhead to reads and writes as both synchronize in the metadata layer. 
Conversely, LV[I] shows lower read overhead and stabilized performance, because reads can now only conflict with extremely outdated table versions. For writes (Figures \ref{fig:YCSB-comp-write-heavy} and \ref{fig:YCSB-comp-read-heavy}), LV[I] shows a larger overhead than LV[R] due to its commit procedure and failed validations requiring full retries. In summary, LV[R] optimizes concurrent writes with faster performance but higher variability, while LV[I] is ideal for stable reads under concurrency without introducing writer's congestion.

\subsection{Benefits and Trade-offs for Transactions}

\begin{figure}
        \centering
        \includegraphics[width=0.45\textwidth]{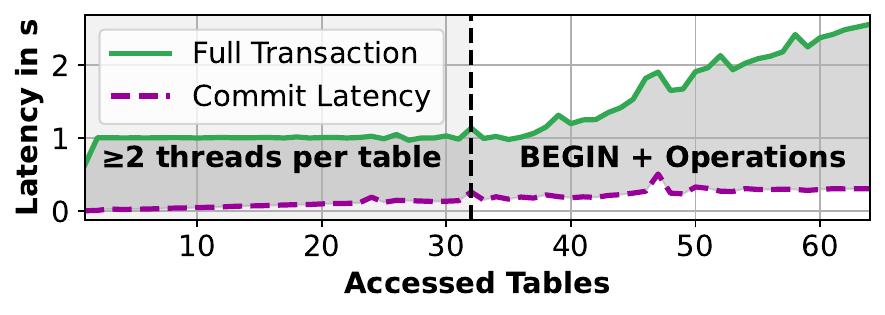}
    \caption{Table Scaling of LV[CT] on MinIO. The client wrote one row to every opened table (matplotlib)}
    \label{fig:cab-lv-table}
\end{figure}

\labeltitle{LV[CT] 's trade-off.}
Next, we evaluate the strengths and performance trade-offs of LV[CT], which supports multiple tables.
Unlike LV[R], LV[CT]'s overhead depends on the number of accessed tables per transaction. We scaled the number of tables on MinIO, following the customer table definition of TPC-H. All transactions inserted a single row into each table to reduce operation overhead but ensure it triggers LV[CT]'s conflict detection across all tables. The LV clients use 64 threads, and we hypothesize that a transaction's commit latency will increase with the number of tables due to more metadata layers and markers being involved.

Our results can be seen in Figure \ref{fig:cab-lv-table}. As expected, we can observe the overhead of LV[CT] 's conflict detection once we access more than one table. 
However, the transaction latency stays stable (at ca. 1s) until we run transactions accessing more than 32 tables. The parallelization in our LV prototype (and assigning 64 threads to our engine) causes this effect. Our prototype uses two threads per table for LV[CT]: one to execute the operations and the other to track dependencies to other tables. Hence, as soon as we run out of threads, some threads will need to manage two or more tables simultaneously, causing increased latency. 

Additionally, the commit latency increases with each accessed table. The tracking of dependencies is the overhead of LV[CT], but this does not impact the overall latency for transactions with less than 32 tables. The expanded commit procedure of generating the log/snapshot entries and overwriting the markers for each accessed table separately impacts this feature minimally.

\begin{figure}
    \centering
 \begin{subfigure}{.45\textwidth}
        \centering
        \includegraphics[width=0.95\textwidth]{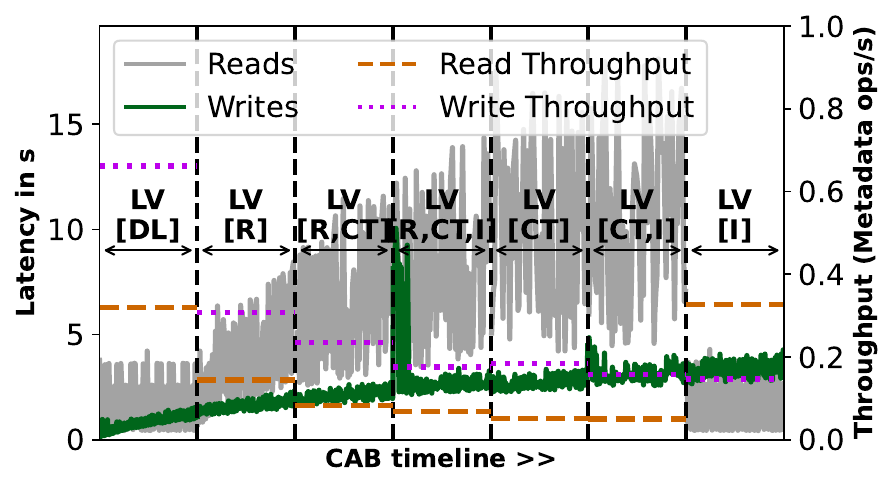}
        \caption{Sequential Runs}
        \label{fig:CAB-reads}
        \label{fig:CAB-writes}
        \label{fig:CAB-seq}
    \end{subfigure}
    \\
    \begin{subfigure}{0.45\textwidth}
        \centering
        \includegraphics[width=0.95\textwidth]{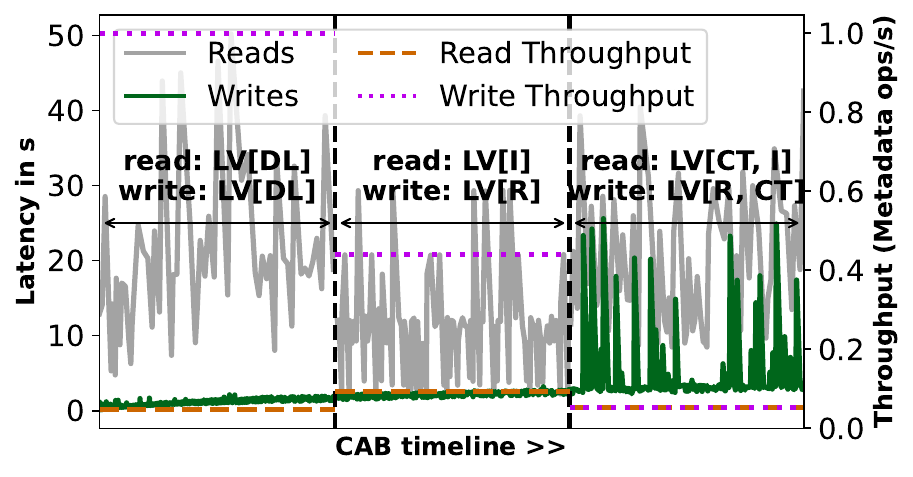}
        \caption{Concurrent reads and writes}
        \label{fig:CAB-concurrent}
    \end{subfigure}
    \caption{TPC-H (S3, sf10) throughput of different LV features of CAB read and write streams \cite{journals/pvldb/RenenL23} (matplotlib) }
    \label{fig:cab-all}
\end{figure}

\labeltitle{CAB-LV: Feature Combinations} 
We now proceed with more complex workloads to investigate the effects on performance when we switch from one LV feature combination to another.
To assess that, we adapted the Cloud Analytics Benchmark (CAB) \cite{journals/pvldb/RenenL23} for LHs and distinguished between read and write streams. Read streams execute the streams generated by the original CAB generator, consisting of TPC-H queries. The write streams either insert up to 100 rows into a random table per query (80\%) or issue update requests on a single table (20\%). Our real-world observations on LH workloads inspired this distribution. We execute the experiment on TPC-H sf10 on S3 and report their latency in Figure \ref{fig:cab-all}. Each phase lasted 20 minutes until a client switches to a new feature combination.

\labeltitle{CAB-LV: Sequential Runs.} Our hypothesis for sequential runs of read and write streams follows our previous observations. Reads will benefit from configurations without added congestion with writes, while writes can use marker-based combinations to minimize aborts.
In Figure \ref{fig:CAB-seq}, we show the latency of sequential executions. We observe the best performance for LV[DL] and LV[I], which avoid the overhead of marker placement.
In contrast, the worst performance is observed when we enable all LV features. LV[CT] 's conflict detection has an overhead when enabled on LV[R] or LV[I]. On its own, LV[CT] has a high latency, but benefits from other features that use markers (compare LV[R, CT] with LV[CT, I]).

\labeltitle{CAB-LV: Recommended Usage.}
We suggest using the original Delta Lake or LV[I] for reads, whereas for writes the decision depends on the desired isolation-level guarantees and the workload itself. Our recommendation is LV[DL] for single-query writes, LV[R] for single-table writes, and LV[R, CT] for multi-table writes.

\labeltitle{CAB-LV: Concurrent Runs.}
We tested our recommendations by executing reads and writes concurrently in our CAB scenario. Figure \ref{fig:CAB-concurrent} shows a plot of the latencies of the metadata operations. In this setting, LV[I] exhibits the lowest latency due to its global version assignment. 
In contrast, with LV[DL] the readers experience a higher overhead because version assignments are done for all tables individually. Concurrent writes using LV[R] are unaffected by concurrent reads with LV[I], and can thus benefit from the marker placement. Effectively, when compared to LV[DL], LV[R] moves the main synchronization to the beginning of a transaction with a minor impact on the performance and stability using its sublog structure. When adding LV[CT] to both operations, we can see more outliers for the writes and slower reads. The readers now need to set markers within the table, making them compete with the writers. However, by using LV[CT], we enable all clients to detect concurrent transactions and react to them (by adapting or requesting a new version) before committing. In summary, we confirm our recommendations, but we also note that the effects were only noticeable because we used a smaller scale factor (sf10). For larger amounts of data, we must account more for data loading, making observable effects minimal.

\subsection{Transactional and Mixed Workloads}

\begin{table}
\centering
\caption{ TPC-C with 1 warehouse and sequential and concurrent clients[successful txns/min] on S3}
\label{tbl:tpccs3sf1}
\begin{tabular}{l|l|ll}
Transaction type& &   sequential  &  concurrent      \\ \hline
single-op& LV[DL]& \textcolor{green}{\faCheck} (4.22)   &  \textcolor{red}{\faClose}      \\ \hline
single-table& LV[R]&  \textcolor{green}{\faCheck} (3.42) &  \textcolor{red}{\faClose}     \\ \hline
\multirow{2}{*}{\makecell{multi-table}}& LV[CT]& \textcolor{green}{\faCheck} (3.49)  & \textcolor{orange}{\faAsterisk} (2.79)    \\
& LV[I] & \textcolor{green}{\faCheck} (3.67)  &   
\textcolor{orange}{\faAsterisk}  (2.81)
\end{tabular}

\small
 \textcolor{green}{\faCheck} = no abots \quad
 \textcolor{orange}{\faAsterisk} = aborts occurred \quad \textcolor{red}{\faClose} = impossible

\end{table}

\labeltitle{Analytical Workloads.}
Further, we analyzed pure OLTP and mixed workloads. To confirm the capabilities of our individual LV features, we tested TPC-C~\cite{tpcc} using one warehouse to increase the occurrence of transactional conflicts. We hypothesize that LV[CT] and LV[I] can execute full TPC-C transactions while showing aborts. In contrast, while working fine when run sequentially, LV[DL] and LV[R] cannot execute TPC-C under concurrency. Table \ref{tbl:tpccs3sf1} shows our results for sequential and concurrent (16 clients) runs, confirming our hypothesis. All configurations were able to execute successfully sequentially. However, LV[DL] and LV[R] could not do a single TPC-C transaction without any consistency violations. LV[CT]'s and LV[I]'s mechanisms successfully detected consistency violations and triggered aborts for erroneous transactions. Although LV[CT] and LV[I] are able to execute full OLTP workloads, they also showcase that LHs generally are not ideal for such scenarios due to their low throughput.

\begin{figure}
    \centering
    \begin{subfigure}{.49\textwidth}
        \centering
        \includegraphics[width=0.95\textwidth]{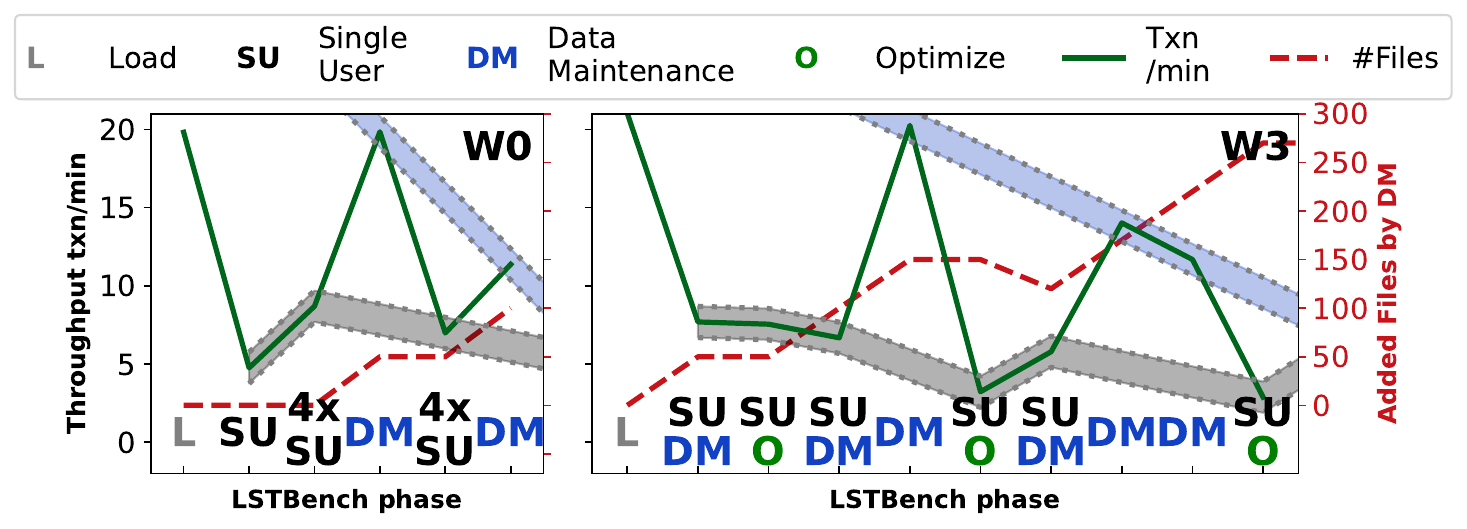}
        \caption{Throughput}
        \label{fig:lstbench-throughput}
    \end{subfigure}\\
    \begin{subfigure}{.49\textwidth}
        \centering
        \includegraphics[width=0.95\textwidth]{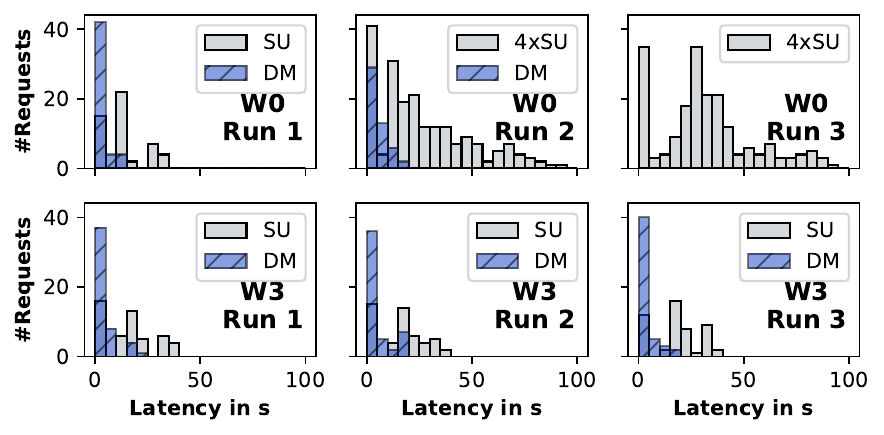}
        \caption{First three repetitions of SU and DM}
        \label{fig:lst-hist}
    \end{subfigure}\\
    \caption{LSTBench (scale factor 100): W0 and W3 with LV[R, CT] (writes) and LV[I] (reads) on S3 (matplotlib)}
    \label{fig:lstbench}
\end{figure}

\begin{figure}
    \centering
        \includegraphics[width=0.45\textwidth]{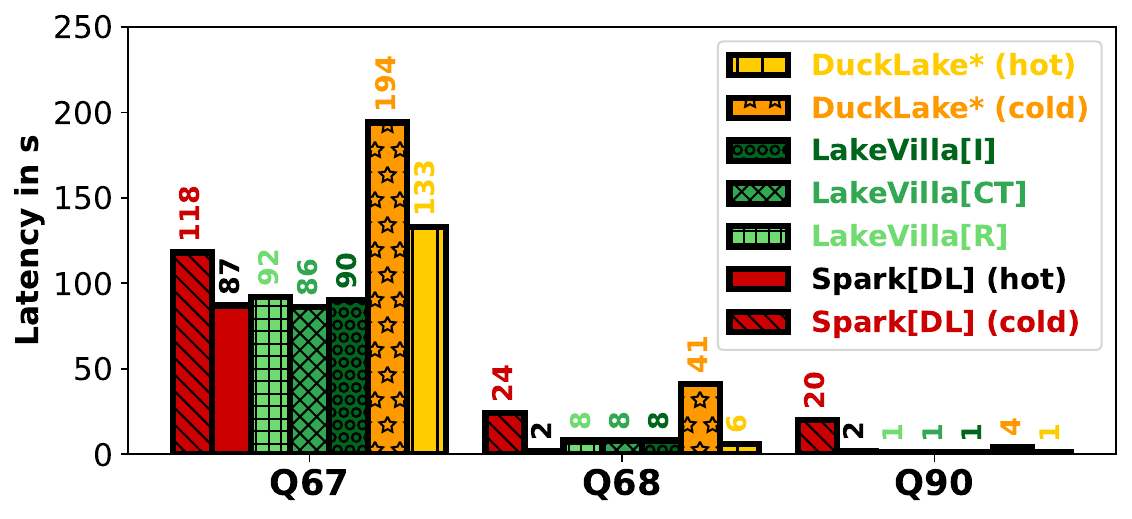}
    \caption{TPC-DS (scale factor 1000): Data loading times of Spark and LakeVilla on MinIO. All entries marked with '*' do not use data partitioning due to technical limitations. (matplotlib)}
    \label{fig:tpcds}
\end{figure}

 \begin{figure}
    \centering
    \begin{subfigure}{.225\textwidth}
        \centering
        \includegraphics[width=0.95\textwidth]{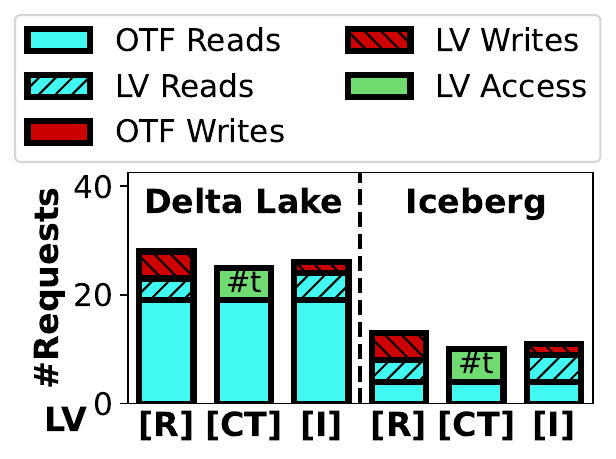}
        \caption{Estimation Spark Read}
        \label{fig:req-read}
    \end{subfigure}%
    \begin{subfigure}{.225\textwidth}
        \centering
        \includegraphics[width=0.95\textwidth]{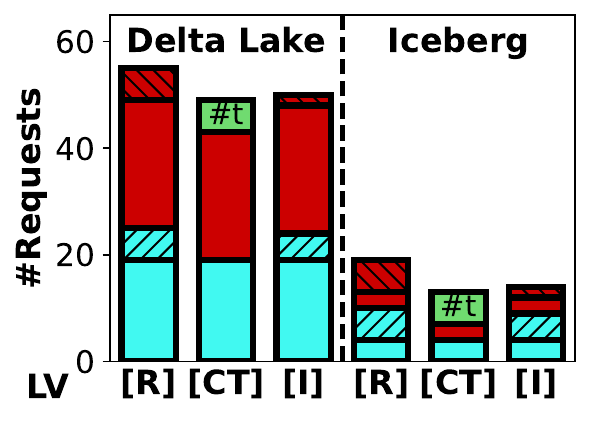}
        \caption{Estimation Spark Write}
        \label{fig:req-write}
    \end{subfigure}\\
    \caption{MinIO statistics: Minimal additional requests needed for LV in Delta Lake and Iceberg (matplotlib)}
    \label{fig:requests}
\end{figure}

\labeltitle{Mixed Workloads.}
After one extreme, we analyzed typical LH scenarios using LSTBench's W0 and W3 \cite{2024lstbench} on TPC-DS scale factor 100 with a selection of 50 random TPC-DS queries (out of Q1-Q10, Q67, Q68, Q90). Following our previous results, we use LV[R, CT] for Data Maintenance (DM), LV[CT, I] for Optimize (O), and LV[I] for SingleUser (SU) tasks. As W0 and W3 focus on concurrent analytics, LV should be able to provide the desired guarantees with minimal impact on performance over a long time.
Figure \ref{fig:lstbench} shows our results when using S3. As theorized, the throughput for respective phases, shown in Figure \ref{fig:lstbench-throughput}, benefits from concurrent SU phases (4xSU in W0) but is impacted by updates introduced by DM phases. O phases compactify the metadata layers for reads in SU, but can not compensate for added data. The compactification effect benefits SU when the metadata is large after multiple updates (compare the first and second O in W3).

Further, when inspecting the latency of the first three occurrences of SU and DM in W0 and W3 in Figure \ref{fig:lst-hist}, we see two types: metadata-intensive (below 1s SUs and DMs) and data-intensive transactions (SU). Metadata-intensive operations primarily interact with the LH metadata layer and stay stable throughout runs. In contrast, data-intensive transactions load extensive data and are impacted by an increasing data size. We can observe this effect with the shift of the largest/second largest bar for SU to the right, the later the W0 run is executed. While LV can keep the Metadata layer at a reasonable size (see O phases), it can not compensate for increased data sizes. We theorize that an integration of LV in analytical systems would improve stability among phases with a built-in cache, as is common for systems interacting with LHs.

\subsection{Analytical workloads}

\labeltitle{Analytics and LakeVilla.}
Lastly, we analyze the impact of LV on pure OLAP using three TPC-DS queries from \cite{conf/cidr/0001KPDSZ23} on scale factor 1000.
These queries rely heavily on OTF metadata design, require selected data, and show the most significant difference across the tested systems \cite{conf/cidr/0001KPDSZ23}.
Our experiment includes LV[R], LV[CT], LV[I], Spark's and DuckLake's cold, and Spark's and DuckLake's best-of-three (hot) performances.
Using the tracing features of MinIO, we measure how long each client interacts with it. 
Additionally, we investigate data caching effects and their influence on OTFs by comparing hot and cold runs. We expect LV's changes in the metadata layer operations will minimally impact performance, as data loading via the network is the main bottleneck.
While LV, as a storage layer, can not fully orchestrate query execution or fully replicate DuckLake's and Spark's task concurrency, our LV results present a lower bound for complete systems. 
Further, we expect noticeable differences between DuckLake's and Spark's hot and cold runs due to avoided object store loads. As DuckLake provides its metadata layer within DuckDB, we expect it to be faster than all alternatives, as it must only load the real data from MinIO.

Our results in Figure \ref{fig:tpcds} mostly confirm our hypothesis. Queries like Q67 primarily depend on data loading, while Q68 and Q90 spend more time on execution-driven or intermediate result caching by Spark and DuckLake. However, DuckLake's cold performance was slower than expected due to the system not partitioning the data on MinIO. It's hot runs for Q68 and Q90 (where it only partially loads the data from MinIO) validate this problem, as they are on par with Spark and LV. However, for Q67 that requires a huge amount of data, DuckLake's cache is unable to compensate for its simplistic data partitioning on the object store, affecting even its hot runs.
Overall, since all LV features show similar performance for all queries, we conclude that LV's features either do not or only exhibit a minor impact on OLAP.

\subsection{Compatibility: LakeVilla + Existing Systems}

\labeltitle{Different Stages and Added Requests.} 
However, one would argue that adding the extra complexity (features) not only induces overhead on the performance but also costs as we perform additional requests to the object store. To investigate, we measured the number of requests throughout our previous experiments and excluded requests not directed to OTF metadata layers. By comparing our baseline LV[DL] with the individual LV features, we determined the required requests for each feature. We expect LV to add a minor number of requests as its design integrates well into mandatory parts of the base OTFs.

Figure \ref{fig:requests} shows the characteristics of Spark's Delta Lake and Iceberg with the added requests for each feature for their single-query transactions. LV's markers and sublog add the most additional write requests. For reads (Figure \ref{fig:req-read}), this impact becomes more apparent as for writes (Figure \ref{fig:req-write}), where Spark already issues such requests. Thus, this further confirms our previous results of LV[R] being more suitable for write transactions. LV[CT] adds a fixed number of one write request and five reads per accessed table. Hence, like its latency, its costs depend on the number of accessed tables within a transaction. Lastly, LV[I] adds the least amount of additional requests, with its added write requests being skippable for read-only transactions in its WriteSerializable mode, integrating it well into read transactions.

\section{Related Work}
\label{sec:related_work}

\labeltitle{Integrating and accessing data lakes.}
Previous research has focused on the operational efficiency of data lakes, like improving latency when accessing files by introducing strategies to encode relational data differently or improve file encodings \cite{journals/pvldb/AfroozehB23,journals/pacmmod/KuschewskiSAL23,conf/cidr/GhitaTB20}. 
Further, Umbra focused on efficient parquet readers \cite{conf/edbt/Rey24,conf/btw/ReyFN23}, transferring required sections of a parquet file and providing insights into the format structure. 
These approaches complement LV because Parquet files are commonly used for LH table data. 
Furthermore, cloud research has also explored elastic computing and additional structures to reduce cost and data lake traffic \cite{journals/pvldb/DurnerL023,journals/pvldb/Kraft0ZBSYZ23,conf/damon/LeeWADBS0KR23,conf/sigmod/0002MA20,conf/sigmod/PerronFDM20,journals/is/Ritter22,conf/nsdi/VuppalapatiMATM20,journals/pvldb/WinterGNK22}. 
These approaches follow the proposed architecture of external locks \cite{journals/pvldb/ArmbrustDPXZ0YM20,conf/cidr/Zaharia0XA21}. In contrast, using a built-in approach for LHs, LV enables multiple features that previously required an auxiliary structure.

\labeltitle{Lakehouses.} Multiple proposed designs enabled missing features \cite{conf/cbd/ChenSLLJ22,conf/icde/DeyFR15,conf/sigmod/LevandoskiCDDEH24,conf/cidr/MaddenDKSCMT22} or specializations~\cite{conf/bigdataconf/BegoliGK21,conf/cidr/Hambardzumyan23} in custom OTFs or other setups. They redesign the metadata layer, introduce file optimizations and orderings, or add auxiliary structures and provide built-in features like security or performance for neural network training. Unlike these, LV is an extension applicable to existing OTFs. Project Nessie implements git-like branching for Apache Iceberg to isolate different transactions (cf. (i) \& (iii)) \cite{project_nessie_web}.
However, Project Nessie is a client catalog, while LV focuses on built-in solutions. Similarly, Photon explores design decisions for a vectorized query engine processing Delta Lake tables like the combined memory management or row- versus column-oriented processing \cite{conf/sigmod/BehmPAACDGHJKLL22}. Photon's insights can become relevant for our LV prototype in the future as both use Delta Lake.

\labeltitle{LV and External Solutions}
Multi-table transaction support is a heavily requested feature for LHs \cite{journals/pvldb/ArmbrustDPXZ0YM20,delta4_multi_table,iceberg_multi_table,iceberg_nessie_doc,ducklake}, making it valuable to compare current advances to LV. The Delta Lake commit owner is a hybrid solution, moving the commit decision to an external service and modifying Delta Lake to guarantee persistence \cite{delta4_multi_table}. In contrast, LV operates on the object store, making it a more independent solution. 
The possible advantages of the commit owners are better commit performance, tailored to Delta Lake, and cleaner metadata layers than LV's markers. 
However, commit owners are presumably limited to a single vendor environment and not adaptive to other OTFs and transaction requirements. 
Additionally, the effects of an out-of-sync commit owner or its ''backfilling'' for compatibility are unknown and require further investigation \cite{delta4_multi_table}.

\labeltitle{DuckLake} moves the Lakehouse metadata layer into DuckDB, using the object store purely for table data \cite{ducklake}. This approach reuses DuckDB features and makes LHs more lightweight \cite{ducklake}. However, we found limitations of this approach while preparing our experiments: First, DuckLake currently does not allow the use of the same LH by multiple sessions and processes inherited by DuckDB's design \cite{duckdb_concurrency}, only allowing sequential transactions. Further, it does not synchronize data changes outside DuckDB or upholds any structure on the object store, making DuckDB a single bottleneck and the data unusable for any other system with access to the object store. Lastly, DuckLake uses no data partitioning on the object store, affecting its performance (see TPC-DS measurements in Section \ref{sec:evaluation}). While DuckLake excels in scalability, it ignores some crucial design aspects seen in previous OTF designs, focused on by LV.

\labeltitle{Future Combinations.}
Hence, we could see a combination of the three used in practice: DuckLake or Commit Owners within a vendor's environment for performance and LV for operations from the outside.

\section{Conclusion}
\label{sec:conclusion}
Current lakehouses provide limited transaction support without auxiliary structures. 
In this work, we introduce LakeVilla to improve transactional models and guarantees of state-of-the-art lakehouses. 
LakeVilla defines three modular features to achieve its goal and provides transactional guarantees up to serializability.
Our results show the trade-offs and compatibility with systems used in practice. In contrast to other solutions in the Lakehouse area, we extend the base open table format on the object store non-invasively and allow transactions to freely choose their guarantees, opening up interesting directions for future architectures and approaches.

\section{Statements and Declarations}

\labeltitle{Funding.} This work was funded by the Chair of Database Systems at TUM and SAP SE. The PhD position of Tobias G\"otz is partially funded by SAP SE. 

\begin{acknowledgements}
We are grateful to the colleagues at the Chair of Database Systems at TUM, the SAP HANA Cloud team, and all members of the SAP HANA Campus for their feedback.

\end{acknowledgements}

\bibliographystyle{spmpsci}     
\bibliography{sample}

\end{document}